\documentclass[superscriptaddress,aip,english,11pt]{revtex4-1}

\usepackage{epsfig}
\usepackage{subfigure}
\usepackage{graphicx}

\usepackage{amsmath}
\usepackage{amssymb}
\usepackage{amsthm}
\usepackage{amsfonts}
\usepackage{amstext}
\usepackage{mathtools}
\usepackage{array} 
\usepackage{cancel}
\usepackage{empheq}
\usepackage{algorithm}
\usepackage{algorithmicx}
\usepackage{algpseudocode}
\usepackage{bm}
\usepackage[mathlines]{lineno}
\usepackage{mathrsfs}

\usepackage[usenames,dvipsnames]{color} 
\definecolor{LightBluishGray}{rgb}{0.95,0.95,1.0}
\definecolor{BlueSky}{RGB}{0,127,255}
\definecolor{shade50}{rgb}{0.5,0.5,0.5}
\definecolor{shade20}{rgb}{0.8,0.8,0.8}
\definecolor{shade10}{rgb}{0.9,0.9,0.9}
\definecolor{shade05}{rgb}{0.95,0.95,0.95}
\definecolor{red50}{rgb}{0.5,1.0,1.0}
\definecolor{blue50}{rgb}{1.0,1.0,0.5}
\usepackage{soul}
\sethlcolor{yellow}

\usepackage[usenames,dvipsnames]{pstricks}
\usepackage{pst-grad} 
\usepackage{pst-plot} 

\usepackage{paralist} 
\usepackage{titlesec}
\usepackage{enumitem}
\setlist{nolistsep}
\usepackage{fancyhdr}
\usepackage[title,titletoc]{appendix}
\usepackage{changepage}

\usepackage[margin=1.0in]{geometry}

\usepackage{babel}
\usepackage{hyperref}
\hypersetup{
  colorlinks   = true, 
  urlcolor     = blue, 
  linkcolor    = blue, 
  citecolor   = red     
}
\usepackage{cleveref}
\crefname{section}{part}{parts}
\crefname{section}{section}{sections}

\newenvironment{frcseries}{\fontfamily{frc}\selectfont}{}

\DeclareFontFamily{OT1}{pzc}{}
\DeclareFontShape{OT1}{pzc}{m}{it}{<-> s * [1.10] pzcmi7t}{}
\DeclareMathAlphabet{\mathpzc}{OT1}{pzc}{m}{it}

\newcommand{\avrg}[1]{\ensuremath{\langle{#1}\rangle}}

\newcommand{\bvec}[1]{\ensuremath{\mathbf{#1}}}
\newcommand{\tvec}[1]{\ensuremath{\tilde{\bvec{#1}}}}

\renewcommand{\bar}[1]{\overline{#1}}

\newcommand{\avom}[1]{\avrg{#1}_{_{\Omega_1,\Omega_2} } }

\begin{document}

\title{Modeling of Inelastic Collisions in a Multifluid Plasma:\\ Ionization and Recombination\footnote{Distribution A. Approved for public release; distribution unlimited}} 

\author{Hai P. Le}
\email{hai.le@ucla.edu}
\affiliation{Department of Mathematics, University of California, Los Angeles, California 90095}
\author{Jean-Luc Cambier}
\thanks{Present address: Air Force Office of Scientific Research, Arlington, Virginia 22203.}
\email{jean\_luc.cambier@us.af.mil}
\affiliation{Air Force Research Laboratory, Edwards AFB, California 93524}


\date{\today}
 
\begin{abstract}
A model for ionization and recombination collisions in a multifluid plasma is formulated using the framework introduced in previous work [{Phys. Plasmas} \textbf{22}, 093512 (2015)]. The exchange source terms for density, momentum and energy are detailed for the case of electron induced ionization and three body recombination collisions with isotropic scattering. The principle of detailed balance is enforced at the microscopic level. We describe how to incorporate the standard collisional-radiative model into the multifluid equations using the current formulation. Numerical solutions of the collisional-radiative rate equations for atomic hydrogen are presented to highlight the impact of the multifluid effect on the kinetics.
\end{abstract}


\maketitle 



\section{Introduction}
The ability to accurately model plasmas in non-local thermodynamic equilibrium (non-LTE) is essential in understanding complex phenomena associated with atomic population kinetics, thermal equilibration and radiation transport. \cite{oxenius_kinetic_1986} Collisional-radiative (CR) models are the most common numerical tool used in simulating non-LTE plasmas; these models are adapted to a wide range of applications ranging from low temperature plasmas to high energy density physics. There have been continuous improvements from theoretical calculations of atomic data and cross sections \cite{bar-shalom_hullac_2001,gu_flexible_2008,zatsarinny_bsr:_2006,jonsson_new_2013,kallman_atomic_2007}, to computational models of time-accurate collisional-radiative kinetics for different plasma regimes\cite{annaloro_vibrational_2014,scott_cretinradiative_2001,chung_flychk:_2005,hansen_hybrid_2007}.  Detailed CR models, however, are very computationally intensive due to the enormous amount of atomic data and elementary cross sections involved in the simulation. Therefore these models are traditionally applicable to problems with low dimensionality or used as a post-processing tool for diagnostics. Recently, multi-dimensional hydrodynamic calculations with CR kinetics have become feasible for moderate size kinetic systems thanks to the recent advances in high performance computing\cite{kapper_ionizing_2011,panesi_collisional_2013}. In addition, many coarse-graining techniques have been developed to further reduce the computational cost associated with modeling the CR kinetics.\cite{le_complexity_2013,munafo_boltzmann_2014,guy_consistent_2015} 

An important issue that must be addressed carefully in the CR modeling process is the treatment of non-thermal populations, e.g., hot electrons from laser produced plasmas or electrons emitted from cathode in a discharge system. A proper treatment of these systems requires solving the kinetic equation for the translational degree of freedom of the particles. The two most common approaches for these types of problem are the ``two-term'' approximation\cite{hagelaar_solving_2005} and Monte Carlo collision method. \cite{longo_monte_2000} These methods however are quite expensive for detailed CR modeling with many atomic states. In previous work\cite{le_modeling_2015}, we propose an alternative approach, which is to use the classical multifluid approximation \cite{braginskii_transport_1965,burgers_flow_1969}, in which non-thermal populations can be treated as separated fluids with their mean velocities and temperatures. The focus of this work is to extend the applicability of the CR models to the multifluid regime. Due to the assumption of individual Maxwellians, the relative drift velocity between two different fluids, if significant, can impact the kinetics of the collisions. Our previous work focuses on the modeling of excitation/deexcitation collisions using the multifluid description. \cite{le_modeling_2015} The significance of the relative drift velocity on the kinetics, hereby referred to as the multifluid effect, is characterized by a nondimensional parameter $\lambda$, which is defined as the ratio of the kinetic versus thermal energies computed using the reduced mass and relative (hydrodynamic) velocities of the two colliding particles. We note that this effect had also been examined by different authors.\cite{burgers_flow_1969,horwitz_ion_1973,barakat_momentum_1981,benilov_momentum_1996,benilov_kinetic_1997,conde_friction-force_2008,le_modeling_2015} In 1969, Burgers presented a framework for deriving exchange source terms for a system of moment equations. \cite{burgers_flow_1969} Although most of his results are for a five-moment system, the framework is rather generic and can be readily applied to other moment systems. Horwitz and Banks derived the momentum and energy exchange rates for charge exchange collisions including the multifluid effect. \cite{horwitz_ion_1973} In their model, deviation from single-fluid results is characterized by a parameter $\delta$, which is essentially the square root of the parameter $\lambda$ defined in our work. Conde et al. studied the friction forces due to Coulomb collision for drifting ions in a partially ionized plasma. \cite{conde_friction-force_2008} Barakat and Schunk \cite{barakat_momentum_1981} derived momentum and energy exchange rates for elastic collisions using various forms of elastic cross sections, e.g., inverse-power interaction, hard sphere and Maxwell molecules. Anisotropic effects are also considered in their work. We remark that all the work described above do not include inelastic and/or reactive collisions. These collisions are briefly considered in Burgers using a simple Bhatnagar-Gross-Krook (BGK) operator. \cite{burgers_flow_1969} A more general model for a reactive collision can be found from the work of Benilov \cite{benilov_momentum_1996,benilov_kinetic_1997}, where the derivation is based on consideration of a general two-body collision of the form $\alpha + \beta \Leftrightarrow \gamma + \delta$. Due to the general description of the collision, the exchange source terms are quite complicated making numerical implementation very challenging.

This paper presents a continuation of our previous work \cite{le_modeling_2015} to the case of ionization and three-body recombination collisions. The modeling of these collisions is more complicated than excitation/deexcitation collisions because they involve more than two particles. Using the mulitifluid approximation, each participating particle (electron, neutral, ion) can be characterized as a fluid with its own set of conservation laws. In the most general case, one can have four different fluids associated with the scattered particle $s$, the target particle $t$, its ionized state $i$ and the free electron $e$. Fortunately, as will be shown, simplifications can be made for the special case of electron induced ionization and recombination, which is of particular interest for most applications. The derivation presented in this work follows naturally from our previous work. Some slight modifications are introduced to avoid complication in mathematical notations.

The rest of the paper is organized as follows. Sec. \ref{sec:transfer} describes the kinematics of the collision. The exchange rates for ionization and recombination collisions are considered in sec. \ref{sec:ion} and \ref{sec:rec} respectively. For ionization collisions, we first formulate the exchange terms for the general case, and then perform a systematic reduction to obtain a set of rate equations applicable for the case of electron induced collisions. For recombination collisions, we only consider the case of electron induced recombination using the same reduction technique. Utilizing these rates, we describe in sec. \ref{sec:CR} how to construct a CR model within the multifluid equations. In Sec. \ref{sec:numerics}, we show the numerical evaluation of the rates, and present zero dimensional calculations to demonstrate the impact of the multifluid effect. Finally, a summary is given in Sec. \ref{sec:conclusion}. Several appendices are also provided to elaborate on the derivation of the exchange rates.

\section{Kinematics}\label{sec:transfer}
Let us consider an inelastic collision between two particles $s$ (scattered) and $t$ (target), the result of which leads to an ionization of $t$ into its ionized stage $i$ and creation of a new electron $e$. The reverse process is a three body recombination collision which involves three particles $s$, $i$ and $e$. Both of these processes can be represented by the following reaction:
\begin{align}
\label{eq:rxn}
s (\bvec{v}_{s_0}) + t (\bvec{v}_{t_0}) \Leftrightarrow s (\bvec{v}_{s_1}) + e (\bvec{v}_{e_2})+ i (\bvec{v}_{i_2})
\end{align}
In the case of an ionization collision, the subscript $0$ denotes pre-collision variables and both subscripts $1$ and $2$ denote post-collision variables. For recombination, we have the reverse order where the subscripts $1$ and $2$ denote pre-collision and $0$ denotes post-collision. These notations are slightly different than the one used in excitation/deexcitation \cite{le_modeling_2015}, but they will prove convenient later in defining the rate coefficients. The species names $s,t,i,e$ also indicate the \textit{fluid} to which the particles belong, hence in the general case we have four different fluids. For $s \equiv e$, we have an electron induced ionization/recombination. Conservations of mass, momentum and energy are expressed as:
\begin{subequations}
\begin{align}
m_t &=  m_i + m_e\\
m_s \bvec{v}_{s_0} + m_t \bvec{v}_{t_0} &= m_s \bvec{v}_{s_1} + m_e \bvec{v}_{e_2} + m_i \bvec{v}_{i_2} \\
\frac{1}{2} m_s \bvec{v}_{s_0}^2 + \frac{1}{2} m_t \bvec{v}_{t_0}^2 &= \frac{1}{2} m_s \bvec{v}_{s_1}^2 + \frac{1}{2} m_e \bvec{v}_{e_2}^2 + \frac{1}{2} m_i \bvec{v}_{i_2}^2 + \varepsilon^*
\end{align}
\end{subequations}
where $\varepsilon^*$ is the ionization energy of the target particle. Let us define the following center-of-mass (COM) and relative velocities for the particles both in the left and right hand side of (\ref{eq:rxn}):
\begin{align}\label{eq:transformation}
\begin{bmatrix}
\bvec{V}_0 \\ 
\bvec{g}_0
\end{bmatrix} 
=
\begin{bmatrix}
\frac{m_s}{M} & \frac{m_t}{M} \\ 
1 & -1
\end{bmatrix} 
\cdot
\begin{bmatrix}
\bvec{v}_{s_0} \\ 
\bvec{v}_{t_0}
\end{bmatrix} ; \qquad
\begin{bmatrix}
\bvec{V}_1 \\ 
\bvec{g}_1 \\
\bvec{g}_2
\end{bmatrix} 
=
\begin{bmatrix}
\frac{m_s}{M} & \frac{m_e}{M} & \frac{m_i}{M}\\ 
1 & -\frac{m_e}{m_t} & -\frac{m_i}{m_t}\\ 
0 & 1 & -1
\end{bmatrix} 
\cdot
\begin{bmatrix}
\bvec{v}_{s_1} \\ 
\bvec{v}_{e_2}  \\
\bvec{v}_{i_2}
\end{bmatrix}
\end{align}
where $M = m_s + m_t = m_s + m_e + m_i$. One can easily verify that the both transformations are unitary, i.e. $d\bvec{V}_0 d\bvec{g}_0 \equiv d\bvec{v}_{s_0} d\bvec{v}_{t_0}$ and $d\bvec{V}_1 d\bvec{g}_1 d\bvec{g}_2 \equiv d\bvec{v}_{s_1} d\bvec{v}_{e_2} d\bvec{v}_{i_2}$. The inverse transformation can be easily found from (\ref{eq:transformation}), leading to:
\begin{align}\label{eq:inverse_transformation}
\begin{bmatrix}
\bvec{v}_{s_0} \\ 
\bvec{v}_{t_0}
\end{bmatrix}
=
\begin{bmatrix}
1 & \frac{m_t}{M} \\ 
1 & -\frac{m_s}{M}
\end{bmatrix} 
\cdot
\begin{bmatrix}
\bvec{V}_0 \\ 
\bvec{g}_0
\end{bmatrix}  ; \qquad
\begin{bmatrix}
\bvec{v}_{s_1} \\ 
\bvec{v}_{e_2}  \\
\bvec{v}_{i_2}
\end{bmatrix}
=
\begin{bmatrix}
1 & \frac{m_t}{M} & 0\\ 
1 & -\frac{m_s}{M} & \frac{m_i}{m_t}\\ 
1 & -\frac{m_s}{M} & -\frac{m_e}{m_t}\\ 
\end{bmatrix} 
\cdot
\begin{bmatrix}
\bvec{V}_1 \\ 
\bvec{g}_1 \\
\bvec{g}_2
\end{bmatrix} 
\end{align}
We can apply the same transformation to the bulk hydrodynamic velocities:
\begin{align}\label{eq:transformation2}
\begin{bmatrix}
\bvec{U}_0 \\ 
\bvec{w}_0
\end{bmatrix} 
=
\begin{bmatrix}
\frac{m_s}{M} & \frac{m_t}{M} \\ 
1 & -1
\end{bmatrix} 
\cdot
\begin{bmatrix}
\bvec{u}_s \\ 
\bvec{u}_t
\end{bmatrix} ; \qquad
\begin{bmatrix}
\bvec{U}_1 \\ 
\bvec{w}_1 \\
\bvec{w}_2
\end{bmatrix} 
=
\begin{bmatrix}
\frac{m_s}{M} & \frac{m_e}{M} & \frac{m_i}{M}\\ 
1 & -\frac{m_e}{m_t} & -\frac{m_i}{m_t}\\ 
0 & 1 & -1
\end{bmatrix} 
\cdot
\begin{bmatrix}
\bvec{u}_{s} \\ 
\bvec{u}_{e}  \\
\bvec{u}_{i}
\end{bmatrix}
\end{align}
Using the COM and relative velocity variables defined in eq. (\ref{eq:transformation}), conservations of momentum and energy can be expressed as:
\begin{subequations}
\begin{align}
M \bvec{V}_0 &= M \bvec{V}_1\\
\frac{1}{2} \mu \bvec{g}_0^2 &=  \frac{1}{2} \mu \bvec{g}_1^2 + \frac{1}{2} \mu_t \bvec{g}_2^2  + \varepsilon^*
\end{align}
\end{subequations}
where $\mu = \frac{m_s m_t}{m_s+m_t}$ and $\mu_t = \frac{m_e m_i}{m_e + m_i}$. Note that conservation of momentum implies that the COM velocity is essentially unchanged after the collision so for simplicity, we can take $\bvec{V} \equiv \bvec{V}_0 = \bvec{V}_1$. Furthermore, let us define $\Upsilon$ to be the energy transferred during the collision:
\begin{align}
\Upsilon &= \frac{1}{2} \mu \bvec{g}_0^2 - \frac{1}{2} \mu \bvec{g}_1^2  =  \frac{1}{2} \mu_t \bvec{g}_2^2 + \varepsilon^*
\end{align}
The last expression is obtained from energy conservation. For the case of ionization/recombination, $\Upsilon \in [\varepsilon^*, \varepsilon]$ where $\varepsilon = \frac{1}{2} \mu \bvec{g}_0^2$ is the available kinetic energy in the COM reference frame. 

\section{Ionization}\label{sec:ion}
\subsection{Transfer integral}
Let us now look at an ionization collision which can be decomposed into a two-step process:
\begin{subequations}\label{eq:ti1}
\begin{align}\label{eq:ti1a}
s (\bvec{v}_{s_0}) + t(\bvec{v}_{t_0}) &\Rightarrow s (\bvec{v}_{s_1}) + t^*(\bvec{v}_{t_1}) \\
\label{eq:ti1b}
t^*(\bvec{v}_{t_1}) &\Rightarrow e (\bvec{v}_{e_2}) + i(\bvec{v}_{i_2})
\end{align} 
\end{subequations}
where the first step is the formation of a \textit{virtual} excited state $t^*$ via scattering and the the second step is a spontaneous ionization of $t^*$. The decomposition of (\ref{eq:ti1}) is used only for the convenience in expressing the exchange variables. We can write a transfer integral expressing the rate of change of any moment variable $\psi$ as follows:
\begin{align}
\Psi^{ion}_{st} = n_s n_t \int d^3 \bvec{v}_{s_0} \,d^3 \bvec{v}_{t_0} \, f_s \, f_t
\, g_0 \int \psi  \omega^{ion}_{st} (\bvec{v}_{s_0},\bvec{v}_{t_0};\bvec{v}_{s_1},\bvec{v}_{e_2},\bvec{v}_{i_2}) \, d^3 \bvec{v}_{s_1} \, d^3 \bvec{v}_{e_2} \, d^3 \bvec{v}_{i_2}
\end{align} 
where $g_0 = | \bvec{g}_0|$ and $\omega^{ion}_{st} (\bvec{v}_{s_0},\bvec{v}_{t_0};\bvec{v}_{s_1},\bvec{v}_{e_2},\bvec{v}_{i_2})$ is the ionization differential cross section. Note that $\Psi^{ion}_{st}$ includes a product of two Maxwellian VDF's $f_s$ and $f_t$. Utilizing the same procedure described in appendix B of Le \& Cambier\cite{le_modeling_2015} for excitation/deexcitation, $\Psi^{ion}_{st}$ can be written in the following form:
\begin{align}
\label{eq:psi_ion_st}
\Psi^{ion}_{st} = n_s n_t \underbrace{  \frac{1}{\pi^\frac{3}{2} a^3} \int d^3 \bvec{V}^*_0 e^{-\bvec{V}^{*2}_0/a^2} }_{\int d^3 \bvec{V}^* f_{V^*}} \cdot 
\frac{1}{\pi^\frac{3}{2} \alpha^3}\int d^3  \bvec{g}_0 \, e^{- \tvec{g}_0^2/\alpha^2} g_0 \int \psi  \omega^{ion}_{st} ( \bvec{g}_0; \bvec{g}_1, \bvec{g}_2) \, d^3  \bvec{g}_1 \, d^3  \bvec{g}_2
\end{align} 
where $\omega^{ion}_{st} ( \bvec{g}_0; \bvec{g}_1, \bvec{g}_2)$ is the differential cross section (DCS) expressed in terms of relative velocities. The average quantities used in the transformation are summarized in table \ref{tab:ion}. Note that these variables are defined only for ionization. For recombination, we have a different set of average variables. Table \ref{tab:ion} also shows the approximation of these average variables for the case of an electron induced ionization by making use of the small mass ratio $m_e/M \ll 1$, and further assuming that $\frac{m_e}{M} \ll \frac{T_e}{T_t}$. The latter assumption is almost always true for most of the practical cases, especially for electron induced collisions with heavy atoms. For brevity, the Boltzmann constant is omitted throughout the text. 
\begin{table}
\centering
\begin{tabular}{|c|c|c|}
\hline 
Variable & definition & e-induced coll. ($s \equiv e$)\\ 
\hline 
$T^*$ & $\frac{M T_s T_t}{m_s T_t + m_t T_s}$ & $T_t$\\
$\tilde{T}$ & $ \frac{m_s T_t + m_t T_s}{M} $ & $T_e$\\
$a$ & $\sqrt{\frac{2T^*}{M}}$ & $\sqrt{\frac{2T_t}{m_t}}$\\ 
$\alpha$ & $\sqrt{\frac{2\tilde{T}}{\mu}}$ & $\sqrt{\frac{2T_e}{m_e}}$\\ 
$\gamma$ & $\frac{\mu}{M}   \frac{T_t-T_s}{\tilde{T}}$ & $\frac{m_e}{M} \frac{T_t - T_e}{T_e}$ \\ 
$\tvec{g}_0$ & $\bvec{g}_0 - \bvec{w}_0$ & \\
$\bvec{V}^*$ & $\bvec{V} - \bvec{U}_0 + \gamma \tvec{g}_0$ &\\
\hline 
\end{tabular} 
\caption{Summary of variables used for ionization. The second column lists the general definition, and the third one is applicable for an electron induced ionization.}\label{tab:ion} 
\end{table}

The DCS can be written as a triply differential cross section (TDCS):
\begin{align}
\omega^{ion}_{st} ( \bvec{g}_0; \bvec{g}_1, \bvec{g}_2) \, d^3  \bvec{g}_1 \, d^3  \bvec{g}_2 &= \frac{d^3 \sigma^{ion}_{st}}{d\Upsilon d\Omega_1 d\Omega_2}  (g_0,\Upsilon,\Omega_1,\Omega_2) \, d \Upsilon \, d\Omega_1 \, d\Omega_2 
\end{align}
where $\Omega_1$ and $\Omega_2$ are the solid angles of $\bvec{g}_1$ and $\bvec{g}_2$.  Also, we can define a singly differential cross section (SDCS) as:
\begin{align}
\frac{d \sigma^{ion}_{st}}{d\Upsilon}  (g_0,\Upsilon) = \int \frac{d^3 \sigma^{ion}_{st}}{d\Upsilon d\Omega_1 d\Omega_2}  \, d\Omega_1 \, d\Omega_2
\end{align}
This can be used as normalization factor to extract the strictly angular-dependent part of the TDCS, from $\mathcal{G}^{ion} = \frac{d^3 \sigma^{ion}_{st}}{d\Upsilon d\Omega_1 d\Omega_2}  /{\frac{d \sigma^{ion}_{st}}{d\Upsilon}}
$ with the normalization $\int \mathcal{G}^{ion} d\Omega_1 d\Omega_2 = 1$. The total ionization cross section can be easily obtained from $\overline{\sigma}^{ion}_{st} = \int \frac{d \sigma^{ion}_{st}}{d\Upsilon} (g_0,\Upsilon) d \Upsilon$. It must be noted that all the cross sections have a threshold being the ionization energy of particle $t$.

Since we are concerned here with the exchanges of density, momentum and energy, the moment variable $\psi$ (scalar or vector) can always be expanded in terms of powers of $\bvec{V}^*$:
\begin{align}
\psi = a + b \bvec{V}^* + c \bvec{V}^{*2} + \cdots
\end{align}
and the expansion is at most quadratic in $\bvec{V}^*$ since we are only considering the exchanges of mass, momentum and energy. Using the fact that $f_{V^*}$ is a Maxwellian, the integration over $\bvec{V}^*$ can be easily performed:
\begin{align}
\int d^3 \bvec{V}^* f_{V^*} = 1; \quad \int d^3 \bvec{V}^* \, \bvec{V}^* \, f_{V^*} = 0; \quad \int d^3 \bvec{V}^* \, \bvec{V}^{*2} \, f_{V^*} = \frac{3}{M} T^*
\end{align}
Therefore all the terms involving $\bvec{V}^*$ can be easily evaluated (or eliminated), leaving us with the terms independent of $\bvec{V}^*$. To evaluate those terms, we consider the following form of the transfer integral:
\begin{align}\label{eq:transfer_ion0}
\Psi^{ion}_{st} = n_s n_t \frac{1}{\pi^\frac{3}{2} \alpha^3} e^{-w_0^2/\alpha^2} \int d^3 \bvec{g}_0 \, e^{-g_0^2/\alpha^2} e^{2 \bvec{g}_0 \cdot \bvec{w}_0/\alpha^2} g_0 \cdot \int \psi  \frac{d^3 \sigma^{ion}_{st}}{d\Upsilon d\Omega_1 d\Omega_2} \, d \Upsilon \, d\Omega_1 \, d\Omega_2
\end{align} 
Without loss of generality, let us choose a coordinated system ($x,y,z$) such that $\bvec{w}_0$ is aligned with the $\bvec{\hat{z}}$ axis. The relative velocities $ \bvec{g}_0$, $ \bvec{g}_1$ and $ \bvec{g}_2$ can be obtained by the following rotations:
\begin{align}\label{eq:coord-ion}
 \bvec{\hat{g}}_0 = R(\varphi, \theta) \cdot \bvec{\hat{w}}_0; \quad
 \bvec{\hat{g}}_1 = R(\phi_1, \chi_1) \cdot  \bvec{\hat{g}}_0; \quad
 \bvec{\hat{g}}_2 = R(\phi_2, \chi_2) \cdot  \bvec{\hat{g}}_0
\end{align}
where the rotation matrix is defined as follows:
\begin{equation}\label{eq:ti7}
R(\varphi,\theta) = \left(\begin{array}{ccc} c_\varphi c_\theta & -s_\varphi & c_\varphi s_\theta \\
                                                                          s_\varphi c_\theta & c_\varphi  & s_\varphi s_\theta \\
                                                                         -s_\theta                &  0             & c_\theta      \end{array}\right)
\end{equation}
Using $d^3  \bvec{g}_0 = g_0^2 d g_0 d \varphi dc_\theta$ where $c_\theta \equiv \cos \theta$, the transfer integral now becomes:
\begin{align}\label{eq:transfer_ion0}
\Psi^{ion}_{st} = n_s n_t \frac{1}{\pi^\frac{3}{2} \alpha^3} e^{-w_0^2/\alpha^2} \int d g_0 \, e^{-g_0^2/\alpha^2} g_0^3 \cdot \int d\varphi dc_\theta  e^{2 g_0 w_0 c_\theta /\alpha^2}  \int \psi  \frac{d^3 \sigma^{ion}_{st}}{d\Upsilon d\Omega_1 d\Omega_2} \, d \Upsilon \, d\Omega_1 \, d\Omega_2
\end{align} 
where $d \Omega_1 = d \phi_1 d c_{\chi_1}$ and $d \Omega_2 = d \phi_2 d c_{\chi_2}$. Let us define an averaging operator as follows:
\begin{align}\label{eq:angular_average}
\avom{ \psi } &= \int  \psi \, \mathcal{G}^{ion} \, d \Omega_1 \, d \Omega_2
\end{align}
Integration over $\varphi$ yields:
\begin{align}
\label{eq:transfer_ion1}
\Psi^{ion}_{st} = n_s n_t \frac{4 \pi}{\pi^\frac{3}{2} \alpha^3} e^{-w_0^2/\alpha^2} \int dg_0 \, e^{-g_0^2/\alpha^2} g_0^3 \cdot \frac{1}{2} \int_{-1}^{1} dc_\theta \, e^{2 g_0 w_0 c_\theta} \cdot \int \avom{\psi}  \frac{d \sigma^{ion}_{st}}{d\Upsilon } (g_0,\Upsilon) d \Upsilon
\end{align} 
We can now define the following normalized energy variables:
\begin{align}
x_0 = \frac{\varepsilon_0}{\tilde{T}} = \frac{\frac{1}{2} \mu  g_0^2}{\tilde{T}}\quad x_1 = \frac{\varepsilon_1}{\tilde{T}} = \frac{\frac{1}{2} \mu  g_1^2}{\tilde{T}}\quad x_2 = \frac{\varepsilon_2}{\tilde{T}} = \frac{\frac{1}{2} \mu_t  g_2^2}{\tilde{T}} \nonumber \\
x^* = \frac{\varepsilon^*}{\tilde{T}} \quad
\upsilon = \frac{\Upsilon}{\tilde{T}} \quad \lambda = \frac{\frac{1}{2}\mu w_0^2}{\tilde{T}}
\end{align}
Using the variables above, we obtain:
\begin{align}
\label{eq:transfer_ion3}
\Psi^{ion}_{st} = n_s n_t \bar{g}_{\tilde{T}} e^{-\lambda} \int_{x^*}^\infty dx_0 \, e^{-x_0} x_0 \cdot \frac{1}{2} \int_{-1}^{1} dc_\theta \, e^{2\sqrt{\lambda x_0} c_\theta} \cdot \int_{x^*}^{x_0} \avom{\psi}  \frac{d \sigma^{ion}_{st}}{d\upsilon } (x_0,\upsilon) d \upsilon
\end{align}
where $\bar{g}_{\tilde{T}} = \sqrt{\frac{8\tilde{T}}{\pi \mu}}$. The exchange rates for moment variables can now be constructed starting from (\ref{eq:transfer_ion1}) or (\ref{eq:transfer_ion3}).

\subsection{Zero$^{\textrm{th}}$-order moment: number density}
\label{sec:zero-mom}
The rate of change of number density due to an ionization collision can be computed by substituting $\psi = 1$ in (\ref{eq:transfer_ion3}). We arrive at the following:
\begin{align}\label{eq:ion_zeroth}
\Gamma^{ion} = n_s n_t \bar{g}_{\tilde{T}} e^{-\lambda} \int_{x^*}^\infty \!dx_0 \, x_0 \,e^{-x_0} \, \zeta^{(0)}(\sqrt{\lambda x_0})  \bar{\sigma}^{ion}_{st}
\end{align} 
where $\zeta^{(0)} (\xi) = \frac{\sinh (2\xi)}{2\xi}$ as defined for the case of excitation/deexcitation.\cite{le_modeling_2015}
Note that $\Gamma^{ion}$ has a very similar form to the case of excitation/deexcitation. In the limit $\lambda \rightarrow 0$, using $\lim_{\xi \rightarrow 0} \zeta^{(0)} (\xi) = 1$, we recover the well-known expression for single-fluid kinetics:
\begin{align}\label{eq:ion_zeroth_sf}
\Gamma^{ion} = n_s n_t \bar{g}_{\tilde{T}} \int_{x^*}^\infty \!dx_0 \, x_0 \,e^{-x_0} \, \bar{\sigma}^{ion}_{st}
\end{align} 
The rate equations for the number densities can be constructed as follows:
\begin{align}
\frac{dn_s}{dt} = 0 ; \quad
\frac{dn_t}{dt} = -\Gamma^{ion} ; \quad
\frac{dn_e}{dt} = +\Gamma^{ion} ; \quad
\frac{dn_i}{dt} = +\Gamma^{ion} 
\end{align} 

\subsection{First-order moment: momentum density}
\label{sec:first-mom}
We first note that for first-order moments, $\psi$ can be represented by a linear combination of $\bvec{V}^*$, $\bvec{g}_p$ $(p=0,1,2)$ and other constant vectors. Since $\left. \Psi^{ion}_{st} \right|_{\psi = \bvec{V}^*} = 0$ as mentioned before, we can neglect all the terms involving $\bvec{V}^*$; the remaining terms can be determined straight forward from the integration. For $\psi = \bvec{g}_p$, the integration results in a vector parallel to the relative drift velocity $\bvec{w}_0$. This is expected from the symmetry of the problem and can also be shown directly from the transfer integral. For convenience, let us define the following friction rate coefficients $R^{ion}$ as follows:
\begin{align}\label{eq:ion_R}
\left. \Psi^{ion}_{st} \right|_{\psi = \mu \bvec{g}_p} = \mu R^{ion}_p \bvec{w}_0; \quad p = 0,1,2
\end{align}
The expressions for these friction coefficients are given in Appendix \ref{app:ion}. We now consider the rate of change of momentum for each particle.

\subsubsection{Scattered particle $s$}
The net rate of momentum exchange of the scattered particle $s$ due to an ionization collision can be determined by substituting $\psi = -m_s (\bvec{v}_{s_0} \!-\! \bvec{v}_{s_1})$ into eq. (\ref{eq:transfer_ion1}), which leads to:
\begin{align}
\bvec{R}^{ion}_{s} = -\frac{4n_s n_t}{\pi^{\frac{1}{2}} \alpha^3} \cdot \int d^3\bvec{V}^* f_{V^*} \cdot \int dg_0 \, g_0^3 \, e^{-g_0^2/\alpha^2}  \cdot \frac{1}{2} \int_{-1}^1 dc_{\theta} \, e^{2g_0 w_0 c_\theta/\alpha^2} \nonumber \\ \int_{x^*}^{x_0} d\upsilon \frac{d{\sigma}^{ion}_{st}}{d\upsilon} \avom{ m_s (\bvec{v}_{s_0} \!-\! \bvec{v}_{s_1})}
\end{align} 
Using $m_s (\bvec{v}_{s_0} - \bvec{v}_{s_1}) = \mu ( \bvec{g}_0 -  \bvec{g}_1)$ and the definitions of the friction coefficients, we can easily express the rate of change of the momentum of fluid $s$ as follows:
\begin{align}
\bvec{R}^{ion}_{s} &= -\mu(R^{ion}_{0}-R^{ion}_{1}) \bvec{w}_0
\end{align}
The full expression can be obtained from the definitions of the coefficients in (\ref{eq:friction_coefs}):
\begin{align}
\bvec{R}^{ion}_{s} &= -\frac{2}{3}\mu \bvec{w}_0 n_s n_t \bar{g}_{\tilde{T}} e^{-\lambda} \int_{x^*}^\infty dx_0 \, x_0^{\frac{3}{2}} \, e^{-x_0} \, \zeta^{(1)} (\sqrt{\lambda x_0}) \int_{x^*}^{x_0} d\upsilon \frac{d{\sigma}^{ion}_{st}}{d\upsilon} \left[ \sqrt{x_0}- \sqrt{x_1} \avom{c_{\chi_1}} \right]
\end{align} 
where $\zeta^{(1)} (\xi) = \frac{3}{4\xi^2} \left[ \cosh (2\xi) - \frac{\sinh (2\xi)}{2\xi} \right]$ and $\lim_{\xi \rightarrow 0} \zeta^{(1)} (\xi) = 1$. Note that the above expression is very similar to the ones for excitation/deexcitation collisions (eq. (38) of Le \& Cambier\cite{le_modeling_2015}).

\subsubsection{Target particles $t$ and $t^*$}
Let us now look at the rates of momentum loss and gain by $t$ and $t^*$ respectively in reaction (\ref{eq:ti1a}). Using (\ref{eq:inverse_transformation}), the pre-collision velocity and momentum of particle $t$ can be expressed as:
\begin{align}
m_t \bvec{v}_{t_0} & = m_t \bvec{V}^* + m_t \bvec{U}_0 -\frac{m_t \gamma}{\mu}  \mu ( \bvec{g}_0-\bvec{w}_0) - \mu  \bvec{g}_0
\end{align}
Similarly, the post-collision momentum of $t^*$ is:
\begin{align}
m_t \bvec{v}_{t_1} & = m_t \bvec{V}^* + m_t \bvec{U}_0 - \frac{m_t \gamma}{\mu} \mu ( \bvec{g}_0-\bvec{w}_0) - \mu  \bvec{g}_1
\end{align}
Using the coefficients defined in (\ref{eq:friction_coefs}) and the identity $\gamma = \frac{\mu}{M} \frac{T_t-T_s}{\tilde{T}}$, we arrive at the following results:
\begin{subequations}\label{eq:R_t}
\begin{align}
\label{eq:R_t0}
\bvec{R}^{ion}_{t} &= -m_t \Gamma^{ion} \bvec{U}_0 - \frac{m_t}{M} \frac{T_t - T_s}{\tilde{T}} \mu(  \Gamma^{ion} - R^{ion}_{0}) \bvec{w}_0 + \mu R^{ion}_{0} \bvec{w}_0 \\
\label{eq:R_tstar}
\bvec{R}^{ion}_{t^*} &= +m_t \Gamma^{ion} \bvec{U}_0 + \frac{m_t}{M} \frac{T_t - T_s}{\tilde{T}} \mu(  \Gamma^{ion} - R^{ion}_{0}) \bvec{w}_0 - \mu R^{ion}_{1} \bvec{w}_0
\end{align}
\end{subequations}
Similar to the previous case, the full expressions can be obtained using the definitions of $\Gamma^{ion}$ and $R^{ion}_p$. The first term on the right hand side of (\ref{eq:R_t0}) or (\ref{eq:R_tstar}) represents the friction due to generation/removal of new particle from the ionization process. These terms also appear in the rate of change of momentum for $s$ as $\pm m_s \Gamma^{ion} \bvec{U}_0$ but the net effect is zero since we assume that particles $s$ before and after the collision belong to the same fluid. The second term describes a thermal friction force since it is proportional to the temperature difference of the reactants. The last term represents the standard friction due to the relative drift of the two fluids $s$ and $t$. One can easily check that 
\begin{align}
\bvec{R}^{ion}_{t} + \bvec{R}^{ion}_{t^*} + \bvec{R}^{ion}_{s} = 0
\end{align}
which is a statement of momentum conservation.

\subsubsection{Electron and ion}
From reaction (\ref{eq:ti1b}), the momentum gain of particle $t^*$ is distributed to the ion and ejected electron. Using the following relations:
\begin{subequations}
\begin{align}
m_e \bvec{v}_{e_2} &= m_e \bvec{v}_{t_1} + \mu_t  \bvec{g}_2 \\
m_i \bvec{v}_{i_2} &= m_i \bvec{v}_{t_1} - \mu_t  \bvec{g}_2
\end{align}
\end{subequations}
The rates of momentum exchange for the ion and ejected electron can be expressed as:
\begin{subequations}\label{eq:R_ei}
\begin{align}
\label{eq:R_e}
\bvec{R}^{ion}_{e} &=  m_e \Gamma^{ion} \bvec{U}_0 + \frac{m_e}{M} \frac{T_t - T_s}{\tilde{T}} \mu(  \Gamma^{ion} - R^{ion}_{0}) \bvec{w}_0 - \frac{m_e}{m_t} \mu R^{ion}_{1} \bvec{w}_0 + \mu_t R^{ion}_{2} \bvec{w}_0 \\
\label{eq:R_i}
\bvec{R}^{ion}_{i} &= m_i \Gamma^{ion} \bvec{U}_0 + \frac{m_i}{M} \frac{T_t - T_s}{\tilde{T}} \mu(  \Gamma^{ion} - R^{ion}_{0}) \bvec{w}_0 - \frac{m_i}{m_t} \mu R^{ion}_{1} \bvec{w}_0 - \mu_t R^{ion}_{2}  \bvec{w}_0
\end{align}
\end{subequations}
The above equations have the same structure as eq. (\ref{eq:R_t}) but with an additional term reflecting the three-body nature of the ionization/recombination processes. Again, one can easily check that momentum conservation is satisfied:
\begin{align}
\bvec{R}^{ion}_{s} + \bvec{R}^{ion}_{t} + \bvec{R}^{ion}_{e} + \bvec{R}^{ion}_{i} = 0
\end{align}

\subsection{Second-order moment: total energy density}
\label{sec:second-mom}
For second-order moment (here we only consider scalar quantities), the exchange variables $\psi$ can be expressed as scalar products of $\bvec{V}^*$, $\bvec{g}_p$ and other constant velocities. We note that since $\int d^3 \bvec{V}^* \, \bvec{V}^* \, f_{V^*} = 0$, all the dot products linear in $\bvec{V^*}$ vanish after the integration. For convenience, let us now define a set of energy transfer coefficients as follows:
\begin{align}\label{eq:ion_J}
\left. \Psi^{ion}_{st} \right|_{\psi =  \bvec{g}_p \cdot  \bvec{g}_q} &= J_{pq}^{ion} \alpha^2 ; \quad p,q = 0,1,2
\end{align}
The explicit expressions for these coefficients are given in Appendix \ref{app:ion}. Note that we also have:
\begin{subequations}\label{eq:ion_J2}
\begin{align}
\left. \Psi^{ion}_{st} \right|_{\psi =  \bvec{g}_p \cdot \bvec{U}_0} &= R_p^{ion} \bvec{w}_0 \cdot \bvec{U}_0 \\
\left. \Psi^{ion}_{st} \right|_{\psi =  \bvec{g}_p \cdot \bvec{w}_0} &= R_p^{ion} \bvec{w}_0^2 = \lambda R_p^{ion} \alpha^2\\
\left. \Psi^{ion}_{st} \right|_{\psi =  \bvec{w}_0 \cdot \bvec{U}_0} &= \Gamma^{ion} \bvec{w}_0 \cdot \bvec{U}_0
\end{align}
\end{subequations}
\subsubsection{Scattered particle $s$}
The rate of change of energy of particle $s$  can be determined from the transfer integral (\ref{eq:transfer_ion1}) by substituting $\psi = \frac{1}{2} m_s (\bvec{v}_{s_0}^2 \!-\! \bvec{v}_{s_1}^2)$:
\begin{align}
Q^{ion}_{s} = -\frac{4n_s n_t}{\pi^{\frac{1}{2}} \alpha^3} \cdot &\int d^3\bvec{V}^* f_{V^*} \cdot \int dg_0 \, g_0^3 \, e^{-g_0^2/\alpha^2} \nonumber \\ &\cdot \frac{1}{2} \int_{-1}^1 dc_{\theta} \, e^{2 g_0 w_0 c_\theta/\alpha^2} \, \int d\upsilon \frac{d{\sigma}^{ion}_{st}}{d\upsilon} \avom{ \frac{1}{2} m_s (\bvec{v}_{s_0}^2 - \bvec{v}_{s_1}^2)}
\end{align} 
The change in the kinetic energy of $s$ can be re-expressed as follows:
\begin{equation}
\begin{split}
\frac{1}{2} m_s (\bvec{v}_{s_1}^2 - \bvec{v}_{s_0}^2) &= \mu ( \bvec{g}_1 -  \bvec{g}_0 ) \cdot \bvec{V} + \frac{m_t}{M} \frac{\mu}{2} ( \bvec{g}_1^2 -  \bvec{g}_0^2 )\\
&= \mu ( \bvec{g}_1 -  \bvec{g}_0 ) \cdot \bvec{V^*} + \mu ( \bvec{g}_1 -  \bvec{g}_0 ) \cdot \bvec{U}_0 + \gamma\mu ( \bvec{g}_0 -  \bvec{g}_1 ) \cdot \tvec{g}_0  - \frac{m_t}{M} \Upsilon
\end{split}
\end{equation}
The integration of the first term is zero since it is linear in $\bvec{V}^*$. The integration of the second term simply yields $\bvec{R}^{ion}_{s} \cdot \bvec{U}_0$. The product in third term can be easily expanded, and the energy transfer rates defined in appendix \ref{app:ion} can be readily used. For the last term, the integration can be carried out using the relation $\Upsilon = \frac{1}{2} \mu \bvec{g}_0^2 - \frac{1}{2} \bvec{g}_1^2$. The total rate of change becomes:
\begin{align}
Q^{ion}_{s} = \bvec{R}^{ion}_{s} \cdot \bvec{U}_0 + \frac{2\mu}{M} (T_t - T_s) \left[ (J^{ion}_{00}-J^{ion}_{01}) - \lambda(R^{ion}_{0} - R^{ion}_{1}) \right]- \frac{m_t}{M} \tilde{T} \left( J^{ion}_{00} - J^{ion}_{11} \right)
\end{align}
This expression is also similar to the one derived for the case of excitation/deexcitation albeit a less compact form (eq. (55) of Le \& Cambier\cite{le_modeling_2015}).

\subsubsection{Target particles $t$ and $t^*$}
The rate of change of the total energies of $t$ and $t^*$ can be determined in a similar fashion. Using (\ref{eq:inverse_transformation}), the kinetic energy of $t$ can be written as:
\begin{align}
\frac{1}{2} m_t \bvec{v}_{t_0}^2 & = \frac{m_t}{M}  \left(  \frac{1}{2} M\bvec{V^*}^2 + \frac{1}{2} M \bvec{U}_0^2  \right) + \frac{m_t \gamma^2}{\mu} \frac{1}{2} \mu \tvec{g}_0^2  + \frac{m_s}{M} \frac{1}{2} \mu  \bvec{g}_0^2\nonumber\\
&  - \frac{m_t}{\mu} \gamma \mu \tvec{g}_0 \cdot \bvec{U}_0 - \mu  \bvec{g}_0 \cdot \bvec{U}_0 + \gamma \mu \tvec{g}_0 \cdot  \bvec{g}_0 + \bvec{V^*} \cdot \left[ \hdots \right]
\end{align}
where we did not explicitly write the terms linear in $\bvec{V}^*$. Substituting the above expression into the transfer integral, we arrive at the following:
\begin{align}
Q^{ion}_{t} =& -\frac{m_t }{M} \Gamma^{ion} \mathcal{E}^* - \frac{m_t \mu}{M^2} \frac{ (T_t - T_s)^2}{\tilde{T}} \left(J^{ion}_{00} - 2\lambda   R^{ion}_{0} + \lambda \Gamma^{ion} \right) - \frac{m_s}{M} \tilde{T} J^{ion}_{00} \nonumber \\
&+ \frac{m_t}{M} \frac{(T_t-T_s)}{\tilde{T}} \mu \left( R^{ion}_{0} -  \Gamma^{ion} \right) \bvec{w}_0 \cdot \bvec{U}_0  + \mu R^{ion}_{0} \bvec{w}_0 \cdot \bvec{U}_0 \nonumber \\
&- \frac{2\mu}{M} (T_t - T_s) \left( J^{ion}_{00} - \lambda R^{ion}_{0} \right)
\end{align}
where $\mathcal{E}^* = \frac{1}{2} M \bvec{U}_0^2 + \frac{3}{2} T^*$ is the total (kinetic + thermal) energies of the COM frame. Note that there are some terms proportional to $(T_t - T_s )^2$; these terms also appear in a general two-body reaction when considering reactants and products as separate fluids (see, for example, Benilov \cite{benilov_momentum_1996}). Similarly for $t^*$, using:
\begin{align}
\frac{1}{2} m_t \bvec{v}_{t_1}^2 & = \frac{m_t}{M}  \left(  \frac{1}{2} M\bvec{V^*}^2 + \frac{1}{2} M \bvec{U}_0^2  \right)  + \frac{m_t \gamma^2}{\mu} \frac{1}{2} \mu \tvec{g}_0^2 + \frac{m_s}{M} \frac{1}{2} \mu  \bvec{g}_1^2 \nonumber \\
& - \frac{m_t}{\mu} \gamma \mu \tvec{g}_0 \cdot \bvec{U}_0 - \mu  \bvec{g}_1 \cdot \bvec{U}_0 + \gamma \mu \tvec{g}_0 \cdot  \bvec{g}_1 + \bvec{V^*} \cdot \left[ \hdots \right]
\end{align}
We arrive at an equivalent expression for the rate of change of total energy of $t^*$:
\begin{align}
Q^{ion}_{t^*} =& \frac{m_t }{M} \Gamma^{ion} \mathcal{E}^* + \frac{m_t \mu}{M^2} \frac{ (T_t - T_s)^2}{\tilde{T}} \left(J^{ion}_{00} - 2\lambda   R^{ion}_{0} + \lambda \Gamma^{ion} \right) + \frac{m_s}{M} \tilde{T} J^{ion}_{11}\nonumber \\
&- \frac{m_t}{M} \frac{(T_t-T_s)}{\tilde{T}} \mu \left( R^{ion}_{0} -  \Gamma^{ion} \right) \bvec{w}_0 \cdot \bvec{U}_0  - \mu R^{ion}_{1} \bvec{w}_0 \cdot \bvec{U}_0 \nonumber \\
&+ \frac{2\mu}{M} (T_t - T_s) \left( J^{ion}_{01} - \lambda R^{ion}_{1} \right)
\end{align}
In the second reaction, this energy is distributed between the ion and the ejected electron.

\subsubsection{Electron and ion}
Using the transformation in appendix \ref{app:threebody}, the kinetic energies of the ion and electron can be expressed as:
\begin{align*}
\bvec{v}_{e_2}^2 &= \bvec{v}_{t_1}^2 + 2\frac{m_i}{m_t} \bvec{v}_{t_1} \cdot  \bvec{g}_2 + \frac{m_i^2}{m_t^2}  \bvec{g}_2^2\\
\bvec{v}_{i_2}^2 &= \bvec{v}_{t_1}^2 - 2\frac{m_e}{m_t} \bvec{v}_{t_1} \cdot  \bvec{g}_2 + \frac{m_e^2}{m_t^2}  \bvec{g}_2^2
\end{align*}
Hence, the kinetic energies are:
\begin{subequations}
\begin{align}
\frac{1}{2} m_e \bvec{v}_{e_2}^2 &= \frac{m_e}{m_t} \frac{1}{2} m_t \bvec{v}_{t_1}^2 + \mu_t \bvec{v}_{t_1} \cdot  \bvec{g}_2 + \frac{m_i}{m_t} \frac{1}{2} \mu_t  \bvec{g}_2^2\\
\frac{1}{2} m_i \bvec{v}_{i_2}^2 &= \frac{m_i}{m_t} \frac{1}{2} m_t \bvec{v}_{t_1}^2 - \mu_t \bvec{v}_{t_1} \cdot  \bvec{g}_2 + \frac{m_e}{m_t} \frac{1}{2} \mu_t  \bvec{g}_2^2
\end{align}
\end{subequations}
Using the rate coefficient defined in (\ref{eq:ion_J}) and (\ref{eq:ion_J2}), we obtain:
\begin{subequations}\label{eq:Qei}
\begin{align}
\label{eq:Qei_a}
Q^{ion}_{e} =& \frac{m_e }{M} \Gamma^{ion} \mathcal{E}^* + \frac{m_e \mu}{M^2} \frac{ (T_t - T_s)^2}{\tilde{T}} \left(J^{ion}_{00}- 2\lambda   R^{ion}_{0} + \lambda \Gamma^{ion} \right) - \frac{m_e}{M} \frac{(T_t-T_s)}{\tilde{T}} \mu \left( R^{ion}_{0} -  \Gamma^{ion} \right) \bvec{w}_0 \cdot \bvec{U}_0 \nonumber \\
&+ \frac{m_s m_e}{M m_t} \tilde{T} J^{ion}_{11} - \frac{ m_e}{m_t} \mu R^{ion}_{1} \bvec{w}_0 \cdot \bvec{U}_0 + \frac{2\mu m_e}{M m_t} (T_t - T_s) \left( J^{ion}_{01} - \lambda R^{ion}_{1} \right) + \mu_t R^{ion}_{2} \bvec{w}_0 \cdot \bvec{U}_0 \nonumber \\
& -\frac{2\mu_t}{M} (T_t - T_s) \left( J^{ion}_{02} - \lambda R^{ion}_{2} \right) - \frac{2m_s \mu_t}{M \mu} \tilde{T} J^{ion}_{12} +  \frac{m_i}{m_t}  \tilde{T} J^{ion}_{22} \\
Q^{ion}_{i} =& \frac{m_i }{M} \Gamma^{ion} \mathcal{E}^* + \frac{m_i \mu}{M^2} \frac{ (T_t - T_s)^2}{\tilde{T}} \left(J^{ion}_{00} - 2\lambda R^{ion}_{0} + \lambda \Gamma^{ion} \right) - \frac{m_i}{M} \frac{(T_t-T_s)}{\tilde{T}} \mu \left( R^{ion}_{0} -  \Gamma^{ion} \right) \bvec{w}_0 \cdot \bvec{U}_0 \nonumber \\
&+ \frac{m_s m_i}{M m_t} \tilde{T} J^{ion}_{11} - \frac{ m_i}{m_t} \mu R^{ion}_{1} \bvec{w}_0 \cdot \bvec{U}_0 + \frac{2\mu}{M} \frac{m_i}{m_t} (T_t - T_s) \left( J^{ion}_{01} - \lambda R^{ion}_{1} \right)- \mu_t R^{ion}_{2} \bvec{w}_0 \cdot \bvec{U}_0 \nonumber \\
& + \frac{2\mu_t}{M} (T_t - T_s) \left( J^{ion}_{02} - \lambda R^{ion}_{2} \right)  + \frac{2m_s \mu_t}{M \mu} \tilde{T} J^{ion}_{12} + \frac{m_e}{m_t} \tilde{T} J^{ion}_{22}
\end{align}
\end{subequations}
It is straight forward to verify that energy conservation is satisfied:
\begin{align}
Q^{ion}_s + Q^{ion}_t + Q^{ion}_e + Q^{ion}_i = \Gamma^{ion} \varepsilon^*
\end{align}

\subsection{Electron induced ionization $t (\bvec{v}_t) + e (\bvec{v}_{e_0}) \Rightarrow e (\bvec{v}_{e_1}) + e (\bvec{v}_{e_2}) + i (\bvec{v}_{i_2})$}
In the previous sections, we derive the exchange terms for a general ionization collision. The resultant equations are rather complicated for practical use. In this section, we perform a systematic reduction of the general system to obtain a set of equations for the special case of an electron induced ionization ($s \equiv e$); this type of collision is relevant for most applications of interest. Taking advantage of the small mass ratio $m_e/M \ll 1$, the following approximations can be used: $\mu \simeq \mu_t \simeq m_e$, $M \simeq m_t \simeq m_i$, and $g_{\tilde{T}} \simeq \bar{v}_e = \sqrt{\frac{8T_e}{\pi m_e}}$. All the average variables are summarized in table \ref{tab:ion} (third column). To further simplify the problem we also assume that the scattering is isotropic, i.e., $\mathcal{G}^{ion} = 1/16 \pi^2$, hence we have $R^{ion}_p = 0$ for $p>0$ and $J^{ion}_{pq} = 0$ for $p \neq q$. 

The reduction proceeds using the following general procedure. We first note that the rates of change of all the moment variables can always be expressed in terms of quantities in the COM frame. These quantities are then distributed to the particles according to some defined mass ratio. Therefore we can reduce the system by taking the limit as $m_e/M \rightarrow 0$ and $m_t/M \rightarrow 1$ for each of the contributed term, that is, each particle (electron or heavy particle) will receive full contribution from terms proportional to $m_t/M$ and none from terms proportional to $m_e/M$. For example, during an ionization collision, the momentum gain/lost from the COM momentum, i.e., $M \bvec{V}$ is only distributed among the target (loss term) and the ion (gain term). 

The rate of change of number densities can be expressed without any simplification:
\begin{align}\label{eq:ei_ionization0}
\frac{dn_e}{dt} = \Gamma^{ion} = -\frac{d n_t}{dt} = \frac{d n_i}{dt}
\end{align}
For rate of change of momentum densities, we can perform the reduction and arrive at the following:
\begin{subequations}\label{eq:ei_ionization_mom}
\begin{align}
\frac{d(\rho_t \bvec{u}_t)}{dt} &= -M \Gamma^{ion} \bvec{U}_0 - \frac{T_t - T_e}{T_e} \mu \mathcal{K}^{ion} \bvec{w}_0 + \mu \mathcal{R}^{ion} \bvec{w}_0  \\
\frac{d(\rho_i \bvec{u}_i)}{dt} &=  +M \Gamma^{ion} \bvec{U}_0 +\frac{T_t - T_e}{T_e} \mu \mathcal{K}^{ion} \bvec{w}_0\\
\frac{d(\rho_e \bvec{u}_e)}{dt} &= - \mu \mathcal{R}^{ion} \bvec{w}_0 
\end{align}
\end{subequations}
where
\begin{align}
\mathcal{R}^{ion} = R^{ion}_{0}; \quad \mathcal{K}^{ion} = \Gamma^{ion} - R^{ion}_{0}
\end{align}
The system of equations above is formally equivalent to the following approximation at the particle level:
\begin{subequations}\label{eq:ei_ionization1}
\begin{align}
m_t \bvec{v}_{t_0} &\simeq M \bvec{V} - \mu \bvec{g}_0  \\
m_i \bvec{v}_{i_2} &\simeq M \bvec{V} - \mu (\bvec{g}_1 + \bvec{g}_2)\\
m_e (\bvec{v}_{e_0} - \bvec{v}_{e_1} - \bvec{v}_{e_2}) &\simeq \mu (\bvec{g}_0  - \bvec{g}_1 - \bvec{g}_2) 
\end{align}
\end{subequations}
It must be noted that all the error terms in (\ref{eq:ei_ionization1}) are $\mathcal{O}(m_e/M)$. For the rate of change of the total energies, we have:
\begin{subequations}\label{eq:ei_ionization_ener}
\begin{align}
\frac{dE_t}{dt} &= -\Gamma^{ion} \mathcal{E}^* - \frac{\mu}{M} \frac{(T_t-T_e)^2}{T_e} \mathcal{W}^{ion} + \frac{T_t - T_e}{T_e} \mu \mathcal{K}^{ion} \bvec{w}_0 \cdot \bvec{U}_0 \nonumber \\
&+ \mu \mathcal{R}^{ion} \bvec{w}_0 \cdot \bvec{U}_0 - \frac{2\mu}{M} (T_t-T_e) \mathcal{J}^{ion} \\
\frac{dE_i}{dt} &= +\Gamma^{ion} \mathcal{E}^* + \frac{\mu}{M} \frac{(T_t-T_e)^2}{T_e} \mathcal{W}^{ion} - \frac{T_t - T_e}{T_e} \mu \mathcal{K}^{ion} \bvec{w}_0 \cdot \bvec{U}_0 \\
\frac{dE_e}{dt} &=-\Gamma^{ion} \varepsilon^* - \mu \mathcal{R}^{ion} \bvec{w}_0 \cdot \bvec{U}_0 + \frac{2\mu}{M} (T_t-T_e) \mathcal{J}^{ion}
\end{align}
\end{subequations}
where
\begin{subequations}
\begin{align}
\mathcal{W}^{ion} &= J^{ion}_{00} - 2\lambda   R^{ion}_{0} + \lambda \Gamma^{ion} \\
\mathcal{J}^{ion} &= J^{ion}_{00} - \lambda   R^{ion}_{0}
\end{align}
\end{subequations}
The system above is equivalent to following approximation at the particle level:
\begin{subequations}\label{eq:ei_ionization2}
\begin{align}
\frac{1}{2} m_t \bvec{v}_{t_0}^2 &\simeq \frac{1}{2} M \bvec{V}^2 - \mu \bvec{V} \cdot \bvec{g}_0  \\
\frac{1}{2} m_i \bvec{v}_{i_2}^2 &\simeq \frac{1}{2} M \bvec{V}^2 - \mu \bvec{V} \cdot ( \bvec{g}_1 + \bvec{g}_2 )   \\
\frac{1}{2} m_e (\bvec{v}_{e_0}^2 - \bvec{v}_{e_1}^2 - \bvec{v}_{e_2}^2) &\simeq \frac{1}{2} \mu (\bvec{g}_0^2 - \bvec{g}_1^2 - \bvec{g}_2^2) + \mu \bvec{V} \cdot ( \bvec{g}_0 - \bvec{g}_1 - \bvec{g}_2 )   \nonumber \\
& = \varepsilon^* + \mu \bvec{V} \cdot ( \bvec{g}_0 - \bvec{g}_1 - \bvec{g}_2 )
\end{align}
\end{subequations}

The system of equations consisting of (\ref{eq:ei_ionization0}), (\ref{eq:ei_ionization_mom}) and (\ref{eq:ei_ionization_ener}) describes the rates of change of number density, momentum and energy for an electron induced ionization collision with isotropic scattering. For numerical calculation, one needs to pre-compute and store three basic rate coefficients $\Gamma^{ion}$, $R^{ion}_0$ and $J^{ion}_{00}$ as functions of $T_e$ and $\lambda$. All the other coefficients $\mathcal{K}^{ion}$, $\mathcal{R}^{ion}$, $\mathcal{J}^{ion}$ and $\mathcal{W}^{ion}$ can be constructed from these basic coefficients. Although not necessary, the isotropic scattering approximation has allowed us to greatly reduce the number of rate coefficients that need to be calculated.

\section{Recombination}\label{sec:rec}
\subsection{Transfer integral}
For recombination, we consider the reverse process of (\ref{eq:ti1}), which involves the following two-step process:
\begin{subequations}
\begin{align}\label{eq:tr1}
e (\bvec{v}_{e_2}) + i(\bvec{v}_{i_2}) &\Rightarrow  t^*(\bvec{v}_{t_1})\\
 s (\bvec{v}_{s_1}) + t^*(\bvec{v}_{t_1}) &\Rightarrow s (\bvec{v}_{s_0}) + t(\bvec{v}_{t_0})
\end{align} 
\end{subequations}
Similar to the case of an ionization collision, we can write a transfer integral as follows:
\begin{align}
\Psi^{rec}_{sei} = n_s n_e n_i \int d^3 \bvec{v}_{s_1} \,d^3 \bvec{v}_{e_2} \,d^3 \bvec{v}_{i_2} \, f_s \, f_e \, f_i
\, g_1 \, g_2 \, \psi \, \omega^{rec}_{sei} (\bvec{v}_{s_1},\bvec{v}_{e_2},\bvec{v}_{i_2};\bvec{v}_{s_0},\bvec{v}_{t_0}) \, d^3 \bvec{v}_{s_0} \, d^3 \bvec{v}_{t_0}
\end{align} 
where $\Psi^{rec}_{sei}$ now contains a product of three Maxwellian distribution functions. In the general case, the three reactants can belong to three different fluids.
\begin{table}
\centering
\begin{tabular}{|c|c|c|}
\hline 
Variable & Definition & e-induced coll. ($s \equiv e$)\\ 
\hline \hline 
$T^*$ & $\frac{MT_s T_e T_i}{m_s T_e T_i + m_e T_s T_i + m_i T_s T_e}$ & $T_i$\\
$\tilde{T}_t$ & $ \frac{m_e T_i + m_i T_e}{m_e+m_i} $ & $T_e$\\
$\tilde{T}$ & $ \frac{m_s T_e T_i + m_e T_s T_i + m_i T_s T_e} {M\tilde{T}_t} $ & $T_e$\\
$\gamma_t$ & $\frac{\mu( T_i - T_e)}{m_e T_i + m_i T_e}$ & $\frac{\mu}{M} \frac{T_i - T_e}{T_e}$\\ 
$\tilde{\delta}$ & $ \frac{m_s T_e T_i}{m_s T_e T_i + m_e T_s T_i + m_i T_s T_e } $ & $\frac{\mu}{M} \frac{T_i}{T_e}$\\
$\tilde{\gamma}$ & $\frac{\mu_t T_s (T_i - T_e)}{m_s T_e T_i + m_e T_s T_i + m_i T_s T_e } $ & $\frac{\mu}{M} \frac{T_i - T_e}{T_e}$\\
$a$ & $\sqrt{\frac{2T^{*}}{M}}$ & $\sqrt{\frac{2T_i}{m_i}}$\\
$\alpha_t$ & $\sqrt{\frac{2\tilde{T}_t}{\mu_t}}$  & $\sqrt{\frac{2T_e}{m_e}}$ \\
$\alpha$ & $\sqrt{\frac{2\tilde{T}}{\mu}}$  & $\sqrt{\frac{2T_e}{m_e}}$\\
$\tvec{g}_p$ & $\bvec{g}_p - \bvec{w}_p$, $p=0,1,2$ &\\
$\bvec{V}^{**}$ & $\bvec{V}-\bvec{U}_1 - \frac{m_s}{M} \tvec{g}_1 +\tilde{\gamma} \tvec{g}_2 + \tilde{\delta} \tvec{g}_1$ &\\
$\bvec{j}$ & $\tvec{g}_1 - \gamma_t \tvec{g}_2$ & \\
$\bvec{m}$ & $\bvec{w}_1 - \gamma_t \bvec{w}_2$ & \\
\hline 
\end{tabular} 
\caption{Summary of variables used for recombination. The second column lists the general definition, and the third one is applicable for an electron-impact three-body recombination.}\label{tab:rec} 
\end{table}
Using the procedure described in appendix \ref{app:threebody}, the transfer integral can be expressed as:
\begin{align}\label{eq:psi_rec_sei}
\Psi^{rec}_{sei} =  n_s n_e n_i \underbrace{ \frac{1}{\pi^\frac{3}{2} a^3 } \int d^3 \bvec{V}^{**} e^{-\bvec{V}^{**2}/a^2} }_{\int d^3 \bvec{V}^{**} f_{V^{**}}} \cdot 
\frac{1}{\pi^\frac{3}{2} \alpha_t^3} \frac{1}{\pi^\frac{3}{2} \alpha^3} \int e^{-\tvec{g}_2^2/\alpha_t^2} \cdot 
\, e^{-(\tvec{g}_1 - \gamma_t \tvec{g}_2)^{2}/\alpha^2} \nonumber \\
\cdot g_1 g_2 \, \psi \, \omega^{rec}_{sei} (\bvec{g_1},\bvec{g}_2;\bvec{g}_0) d^3 \bvec{g}_0 d^3 \bvec{g}_1  d^3 \bvec{g}_2
\end{align} 
where all the average quantities are listed in table \ref{tab:rec}. Similar to the case of an ionization collision, the integration over $\bvec{V}^{**}$ can be easily eliminated since $f_{V^{**}}$ is a Maxwellian. Therefore we only need to consider the case where $\psi$ is independent of $\bvec{V}^{**}$. The transfer integral can be arranged into:
\begin{align}
\Psi^{rec}_{sei} = \frac{n_s n_e n_i}{\pi^3 \alpha_t^3 \alpha^3}  \Lambda \int F_1 \, F_2 \cdot g_1 g_2 \, \psi \, \omega^{rec}_{sei} (\bvec{g}_1,\bvec{g}_2;\bvec{g}_0) d^3 \bvec{g}_0 d^3 \bvec{g}_1  d^3 \bvec{g}_2
\end{align} 
where the product of all the exponential terms is separated into three parts:
\begin{subequations}
\begin{align}
\Lambda &= e^{-w_2^2/\alpha_t^2} e^{-m^2/\alpha^2}\\
F_1 &= e^{-g_2^2/\alpha_t^2} \cdot e^{-g_1^2/\alpha^2} \cdot e^{-\gamma_t^2 g_2^2/\alpha^2}\\
F_2 &= e^{2 \bvec{g}_2 \cdot \bvec{w}_2 / \alpha_t^2} \cdot e^{2 \gamma_t \bvec{g}_1 \cdot \bvec{g}_2 / \alpha^2} \cdot e^{2 \bvec{g}_1 \cdot \bvec{m} / \alpha^2} \cdot e^{-2 \gamma_t \bvec{g}_2 \cdot \bvec{m} / \alpha^2}
\end{align}
\end{subequations}
For given values of mean velocities and temperatures, $\Lambda$ is fixed, $F_1$ is angular-independent, and $F_2$ is angular-dependent. It is more convenient to introduce the detailed balance (DB) relation aka Fowler relation \cite{oxenius_kinetic_1986} at this point:
\begin{align}
g_1 g_2  \, \omega^{rec}_{sei} (\bvec{g}_1,\bvec{g}_2;\bvec{g}_0) = \frac{\mathpzc{g}_t}{2 \mathpzc{g}_i} \frac{h^3}{\mu_t^3} g_0 \, \omega^{ion}_{st}(\bvec{g}_0;\bvec{g}_1,\bvec{g}_2)
\end{align}
where $\mathpzc{g}$ is the degeneracy weight of the atomic state and $h$ is the Planck constant. Substituting the DB relation back to the transfer integral, we obtain:
\begin{align}\label{eq:transfer_rec0}
\Psi^{rec}_{sie} = \frac{\mathpzc{g}_n}{2 \mathpzc{g}_i \mathbb{Z}_t} \frac{n_s n_i n_e}{\pi^{3/2} \alpha^3} \Lambda \int F_1 \cdot F_2 \cdot g_0 \, \psi \, \omega^{ion}_{st}(\bvec{g}_0;\bvec{g}_1,\bvec{g}_2) d^3 \bvec{g}_0 d^3 \bvec{g}_1  d^3 \bvec{g}_2
\end{align} 
where $\mathbb{Z}_t \equiv \frac{(2\pi \mu_t \tilde{T}_t)^{3/2}}{h^3}$ is the translational partition function defined using the reduced mass and temperature of particle $t$. We can see that the integrand of $\Psi^{rec}_{sie}$ is very similar to the one in (\ref{eq:transfer_ion0}) for ionization but with different exponential weighting functions. Note that $F_1$ and $F_2$ contain terms which are dependent on $\bvec{g}_1$ and $\bvec{g}_2$, so they must be integrated together with the differential cross section.

To proceed, let us define a reference frame $(x,y,z)$ such that $\bvec{m}$ is aligned with the $\bvec{\hat{z}}$ axis. The remaining velocity vectors $\hat{\bvec{w}}_2$, $\hat{\bvec{g}}_0$, $\hat{\bvec{g}}_1$ and $\hat{\bvec{g}}_2$ can be defined according to the following rotation operations:
\begin{align}
\bvec{\hat{w}}_2 &= R(\varphi_w,\theta_w) \cdot \bvec{\hat{m}};\quad
\bvec{\hat{g}}_0 &= R(\varphi,\theta) \cdot \bvec{\hat{m}};\quad
\bvec{\hat{g}}_1 &= R(\phi_1,\chi_1) \cdot \bvec{\hat{g}}_0;\quad
\bvec{\hat{g}}_2 &= R(\phi_2,\chi_2) \cdot \bvec{\hat{g}}_0
\end{align}
where $\varphi_w$ and $\theta_w$ are fixed. Note that this choice of the coordinate system is not unique. In the rotated frame $(\xi,\eta,\varsigma)$ where $\bvec{\hat{g}}_0$ is aligned with $\hat{\varsigma}$, the dot products in $F_2$ can be expanded as:
\begin{subequations}
\begin{align}
\bvec{\hat{g}}_1 \cdot \bvec{\hat{m}} &=  c_\theta c_{\chi_1} - s_\theta s_{\chi_1} c_{\phi_1} \\
\bvec{\hat{g}}_2 \cdot \bvec{\hat{m}} &=  c_\theta c_{\chi_2} - s_\theta s_{\chi_2} c_{\phi_2} \\
\bvec{\hat{g}}_1 \cdot \bvec{\hat{g}}_2 &= c_{\chi_1} c_{\chi_2} + s_{\chi_1} s_{\chi_2} c_{\phi_1 - \phi_2}\\
\bvec{\hat{g}}_2 \cdot \bvec{\hat{w}}_2 &=  f(\varphi_w,\theta_w,\varphi,\theta,\phi_2,\chi_2)
\end{align}
\end{subequations}
For reason of brevity, we did not write the explicit expression for $f$. Using the same averaging operator defined in (\ref{eq:angular_average}), the transfer integral can be rewritten as:
\begin{align}\label{eq:transfer_rec1}
\Psi^{rec}_{sie} = \frac{\mathpzc{g}_n}{2 \mathpzc{g}_i \mathbb{Z}_t}  \frac{n_s n_i n_e}{\pi^{3/2} \alpha^3}  \Lambda \int dg_0 \, g_0^3 \, \int d \varphi dc_\theta \, \int F_1\, \avom{ F_2 \, \psi } \,  \frac{d\sigma^{ion}_{st}}{d\Upsilon} d \Upsilon
\end{align} 
From conservation of energy, $F_1$ can be rewritten as:
\begin{align}
F_1 &= e^{-g^2_2/\alpha_t^2} \cdot \, e^{-g^2_1/\alpha^2} \cdot \, e^{-\gamma_t^2 g^2_2/\alpha^2}\nonumber \\
& = e^{\xi \varepsilon^*/\tilde{T}} e^{-\varepsilon_0/\tilde{T}} e^{(1-\xi)\Upsilon/\tilde{T}}
\end{align}
where $\xi = \frac{\tilde{T}}{\tilde{T}_t} + \gamma_t^2 \frac{\mu}{\mu_t}$. Using nondimensional energy variables, the transfer integral becomes:
\begin{align}\label{eq:transfer_rec3}
\Psi^{rec}_{sie} = \frac{\mathpzc{g}_n}{2 \mathpzc{g}_i \mathbb{Z}_t} n_s n_i n_e \frac{\bar{g}_{\tilde{T}}}{4 \pi} \Lambda e^{\xi x^*} \int_{x^*}^\infty dx_0 \, e^{-x_0}  \cdot x_0 \cdot \int d \varphi dc_\theta \, \int_{x^*}^{x_0} \, e^{(1-\xi)\upsilon} \avom{F_2 \, \psi} \, \frac{d\sigma^{ion}_{st}}{d\upsilon} d\upsilon
\end{align}
Note that the above expression is the most general form of the transfer integral for a recombination collision, and various exchange source terms can be constructed in a similar manner. However, one can see that the rates need to be parametrized in terms of $\tilde{T},\tilde{T}_t,\gamma_t,\lambda_1,\lambda_2,\varphi_w,\theta_w$ where $\lambda_1 = \frac{\bvec{w}_1^2}{\alpha^2}$ and $\lambda_2 = \frac{\bvec{w}_2^2}{\alpha_t^2}$. This is clearly not realistic for any numerical calculation due to excessive storage requirement. Therefore, in this work we will only consider the special case of an electron induced recombination, which allows us to make further assumptions to simplify the description of the exchange coefficients.

\subsection{Electron induced recombination $e (\bvec{v}_{e_1}) + e (\bvec{v}_{e_2}) + i (\bvec{v}_{i_2}) \Rightarrow t (\bvec{v}_t) + e (\bvec{v}_{e_0})$}
Let us now examine the case of an electron induced recombination with isotropic scattering, i.e., $\mathcal{G} = 1/16 \pi^2$. Due to the small mass ratio $m_e/M \ll 1$, the average quantities can be approximated as listed in table \ref{tab:rec} (third collumn). In addition, we also have: $\mu \simeq \mu_t \simeq m_e$, $M \simeq m_t \simeq m_i$, $\mathbb{Z}_t \simeq \mathbb{Z}_e$, $\lambda = \lambda_1 \simeq \lambda_2$, $\varphi_w \simeq \theta_w \simeq 0$ and $\xi \simeq 1+\gamma_t^2$. Here we also assume that $\frac{m_e}{m_i} \ll \frac{T_e}{T_i}$ such that $\gamma_t \ll 1$. As mentioned before, this assumption holds for a wide range of physical domains of interest. Hereafter the subscripts in the differential cross sections denoting colliding partners are omitted for brevity. The transfer integral (\ref{eq:transfer_rec1}) becomes:
\begin{align}\label{eq:eir}
\Psi^{rec}_{eie} = \frac{\mathpzc{g}_n}{2 \mathpzc{g}_i \mathbb{Z}_e}  \frac{ n_e n_i n_e}{\pi^{3/2} \alpha^3}  \Lambda \int dg_0 \, g_0^3 \, \int d \varphi dc_\theta \int F_1\, \avom{ F_2 \psi } \,  \frac{d\sigma^{ion}}{d\Upsilon} d \Upsilon
\end{align}
Using the definitions of $\tilde{\delta}$ and $\tilde{\gamma}$ in (\ref{eq:e20}), we also have:
\begin{align}
\tilde{\delta} \simeq \frac{\mu}{M} \frac{T_i}{T_e}; \quad
\tilde{\gamma} \simeq \gamma_t \simeq \frac{\mu}{M} \frac{T_i-T_e}{T_e} \simeq \tilde{\delta} - \frac{\mu}{M}
\end{align}
The product of the exponential terms can be approximated as:
\begin{subequations}\label{eq:F}
\begin{align}
\Lambda & \simeq e^{-2 w_1^2/\alpha^2 } \\
F_1 & \simeq e^{ \varepsilon^*/\tilde{T}} e^{-\varepsilon_0/\tilde{T}}\\
\label{eq:E2}
F_2 &\simeq  e^{2 \bvec{g}_1 \cdot \bvec{w}_1/\alpha^2} e^{2 \bvec{g}_2 \cdot \bvec{w}_1/\alpha^2}
\end{align}
\end{subequations}
Note that we have neglected terms of $\mathcal{O} (\gamma_t)$ and higher in (\ref{eq:F}); these terms correspond to thermal nonequilibrium effect between the ion and electrons. However this effect is weaker than the multifluid effect (note the multiplication of the mass ratio of $m_e/m_i$ in the definition of $\gamma_t$ and $\tilde{\delta}$). Hence the assumptions in (\ref{eq:F}) are reasonable for a wide range of conditions. These approximations are equivalent to neglecting terms of $\mathcal{O} (\gamma_t)$ directly from eq. (\ref{eq:psi_rec_sei}), i.e., $\bvec{j} \simeq \tvec{g}_1$ and $\bvec{m} \simeq \bvec{w}_1$. We have also performed the integration of the full transfer integral (\ref{eq:eir}) and the results indicate that the rates are very weakly dependent on $\gamma_t$. The errors due to the approximations in (\ref{eq:F}) are negligible, with some discrepancies observed only for the case of $\psi = \bvec{g}_1 \cdot \bvec{g}_2$. However, the errors are not very significant and only limited to the region of large $\gamma_t$ ($T_i \gg T_e$), which again falls outside of our physical domain of interest. Nevertheless, these approximations allow us to reduce the parameter space to characterize the exchange rates, and obtain a more compact form of the transfer integral. 

For the case of isotropic scattering, it is more convenient to define a LAB reference frame such that $\bvec{w}_1$ is aligned with the $\bvec{\hat{z}}$ axis and rotated frames such that $\bvec{\hat{g}}_0 = R(\varphi,\theta) \cdot \bvec{\hat{w}}_1$, $\bvec{\hat{g}}_1 = R(\phi_1,\chi_1) \cdot \bvec{\hat{w}}_1$ and $\bvec{\hat{g}}_2 = R(\phi_2,\chi_2) \cdot \bvec{\hat{w}}_1$. $F_2$ then becomes:
\begin{align}
F_2 =e^{2 g_1 w_1 c_{\chi_1}/\alpha^2} e^{2 g_2 w_1 c_{\chi_2}/\alpha^2} 
=e^{2 \sqrt{\lambda x_1} c_{\chi_1} } e^{2 \sqrt{\lambda x_2} c_{\chi_2}}
\end{align}
Using non-dimensional energy variables and after a trivial integration over $\varphi$ and $c_\theta$, the transfer integral is:
\begin{align}\label{eq:eir2}
\Psi^{rec}_{eie} = \frac{\mathpzc{g}_n}{2 \mathpzc{g}_i \mathbb{Z}_e}  n_e n_i n_e \bar{g}_{\tilde{T}} e^{-2 \lambda } e^{x^*} \int_{x^*}^\infty dx_0 \, e^{-x_0} \cdot x_0 \, \int_{x^*}^{x_0}  \avom{ F_2 \psi } \,  \frac{d\sigma^{ion}}{d\upsilon} d \upsilon
\end{align}

\subsection{Zero$^{\textrm{th}}$-order moment: number density}
For zeroth order exchange rate, substituting $\psi = 1$ into (\ref{eq:eir2}) leads to:
\begin{align}
\label{eq:rec_Gamma}
\Gamma^{rec} = \frac{\mathpzc{g}_n}{2 \mathpzc{g}_i \mathbb{Z}_e} n_i n_e^2 \bar{g}_{\tilde{T}} e^{-2 \lambda } e^{ x^*} \int_{x^*}^\infty dx_0 \, e^{-x_0} \cdot x_0 \, \int_{x^*}^{x_0}  \zeta^{(0)} \left( \sqrt{\lambda x_1} \right) \zeta^{(0)} \left( \sqrt{\lambda x_2} \right) \frac{d\sigma^{ion}}{d\upsilon} d \upsilon
\end{align}
where $\zeta^{(0)} (\xi)$ is defined the same as before. One can easily check that in the limit of $\lambda \rightarrow 0$, we recover the Saha equation:
\begin{align}
\lim_{\lambda \rightarrow 0} \frac{\varpi^{rec}}{\varpi^{ion}} = \frac{\mathpzc{g}_n}{2 \mathpzc{g}_i \mathbb{Z}_e} e^{ x^*} 
\end{align}
where $\varpi^{ion} \equiv \Gamma^{ion}/n_t n_e$ and $\varpi^{rec} \equiv \Gamma^{rec}/n_i n_e^2$ are the ionization and recombination rates. Note that the parameter $\lambda$ is defined differently for ionization and recombination. 

The rate equations for the number densities due to recombination can be constructed as follows:
\begin{align}
\frac{dn_t}{dt} = +\Gamma^{rec}; \quad
\frac{dn_e}{dt} = -\Gamma^{rec}; \quad
\frac{dn_i}{dt} = -\Gamma^{rec} 
\end{align} 

\subsection{First-order moment: momentum density}
Similar to the case of ionization, the integral with $\psi = \bvec{g}_p$ results in a vector proportional to the relative drift velocity $\bvec{w}_1$. Let us define the following friction coefficients for recombination:
\begin{align}\label{eq:rec_R}
\left. \Psi^{rec}_{eie} \right|_{\psi = \mu \bvec{g}_p} = \mu R^{rec}_p \bvec{w}_1; \quad p = 0,1,2
\end{align}
The explicit forms of these coefficients are given in (\ref{eq:friction_coefs_rec}). In order to compute the exchange rates for momentum densities, we can start from the approximation in (\ref{eq:ei_ionization1}), and arrive at:
\begin{subequations}
\begin{align}
m_t \bvec{v}_{t_0} 
& \simeq M \bvec{V}^{**} + M \bvec{U}_1 -\tilde{\gamma} M (\tilde{\bvec{g}}_1 + \tilde{\bvec{g}}_2) - \mu  \bvec{g}_0 \\
m_i \bvec{v}_{i_2} &\simeq  M\bvec{V}^{**} + M\bvec{U}_1 - \tilde{\gamma} M (\tilde{\bvec{g}}_1 + \tilde{\bvec{g}}_2) - \mu ( \bvec{g}_1 + \bvec{g}_2)\\
m_e (\bvec{v}_{e_0} - \bvec{v}_{e_1} - \bvec{v}_{e_2}) &\simeq  \mu (\bvec{g}_0 - \bvec{g}_1 - \bvec{g}_2 )
\end{align}
\end{subequations}
Substituting these expressions for the exchange variables, we obtain:
\begin{subequations}
\begin{align}
\bvec{R}_t^{rec} & = +M \Gamma^{rec} \bvec{U}_1 + \frac{T_i - T_e}{T_e}  \mu (2\Gamma^{rec} - R^{rec}_1 - R^{rec}_2) \bvec{w}_1 - \mu  R^{rec}_0 \bvec{w}_1\\
\bvec{R}_i^{rec} & = -M \Gamma^{rec} \bvec{U}_1 - \frac{T_i - T_e}{T_e}  \mu (2\Gamma^{rec} - R^{rec}_1 - R^{rec}_2) \bvec{w}_1 + \mu  (R^{rec}_1 + R^{rec}_2) \bvec{w}_1 \\
\bvec{R}_e^{rec} & = \mu (R^{rec}_0 - R^{rec}_1 - R^{rec}_2) \bvec{w}_1
\end{align}
\end{subequations}
For isotropic scattering, it is easy to see that $R^{rec}_0 = 0$ so we can re-write the above equations into the same form as (\ref{eq:ei_ionization_mom}):
\begin{subequations}
\begin{align}
\bvec{R}_t^{rec} & = +M \Gamma^{rec} \bvec{U}_1 + \frac{T_i - T_e}{T_e}  \mu \mathcal{K}^{rec} \bvec{w}_1 \\
\bvec{R}_i^{rec} & = -M \Gamma^{rec} \bvec{U}_1 - \frac{T_i - T_e}{T_e}  \mu \mathcal{K}^{rec} \bvec{w}_1 + \mu  \mathcal{R}^{rec} \bvec{w}_1 \\
\bvec{R}_e^{rec} & = -\mu \mathcal{R}^{rec} \bvec{w}_1
\end{align}
\end{subequations}
where
\begin{subequations}
\begin{align}
\mathcal{R}^{rec} &= R^{rec}_1 + R^{rec}_2 \\
\mathcal{K}^{rec} &= 2\Gamma^{rec} - R^{rec}_1 - R^{rec}_2
\end{align}
\end{subequations}

\subsection{Second-order moment: energy density}
For second order moment, we can define a set energy exchange coefficients for recombination:
\begin{align}\label{eq:rec_J}
\left. \Psi^{rec}_{eie} \right|_{\psi =  \bvec{g}_p \cdot  \bvec{g}_q} &= J_{pq}^{rec} \alpha^2 ; \quad p,q = 0,1,2
\end{align}
The explicit forms of these coefficients are given in (\ref{eq:thermal_coefs_rec}). We can use the same approximation in (\ref{eq:ei_ionization2}) to express the kinetic energy of each particle in terms of variables in the COM frame. The total kinetic energy of the COM motion $\frac{1}{2} M \bvec{V}^2$ can be expressed as:
\begin{align}
\frac{1}{2} M \bvec{V}^2 = \frac{1}{2} M \bvec{V}^{**2} + \frac{1}{2} M \bvec{U}_1^2 + \tilde{\gamma}^2 \frac{1}{2} M (\tvec{g}_1 + \tvec{g}_2)^2 - \tilde{\gamma} M \bvec{U}_1 \cdot (\tvec{g}_1 + \tvec{g}_2) + \bvec{V}^{**}\cdot \left[ \hdots \right] 
\end{align}
Therefore, the rate equations for energy densities can be written as:
\begin{subequations}\label{eq:ei_rec1b}
\begin{align}
\label{eq:ei_rect1b}
Q_t &= +\Gamma^{rec} \mathcal{E}^* + \frac{\mu}{M} \frac{(T_i - T_e)^2}{T_e} \mathcal{W}^{rec} + \frac{T_i - T_e}{T_e} \mu \mathcal{K}^{rec} \bvec{w}_1 \cdot \bvec{U}_1  \\
\label{eq:ei_reci1b}
Q_i &= -\Gamma^{rec} \mathcal{E}^* - \frac{\mu}{M} \frac{(T_i - T_e)^2}{T_e} \mathcal{W}^{rec} - \frac{T_i - T_e}{T_e} \mu \mathcal{K}^{rec} \bvec{w}_1 \cdot \bvec{U}_1 \nonumber \\
 &+ \mu \mathcal{R}^{rec} \bvec{w}_1 \cdot \bvec{U}_1 - \frac{2\mu}{M} (T_i - T_e) \mathcal{J}^{rec} \\
Q_e & = \Gamma^{rec} \varepsilon^* - \mu \mathcal{R}^{rec} \bvec{w}_1 \cdot \bvec{U}_1 + \frac{2\mu}{M} (T_i - T_e) \mathcal{J}^{rec}
\end{align}
\end{subequations}
where $\mathcal{E}^* = \frac{3}{2} T^{*} + \frac{1}{2} M \bvec{U}_1^2$ and
\begin{subequations}
\begin{align}
\mathcal{W}^{rec} &= J^{rec}_{11} + J^{rec}_{22} + 2J^{rec}_{12} + 4\lambda \Gamma^{rec} - 4\lambda R^{rec}_1 - 4\lambda R^{rec}_2  \\
\mathcal{J}^{rec} &= J^{rec}_{11} + J^{rec}_{22} + 2J^{rec}_{12} - 2\lambda R^{rec}_1 - 2\lambda R^{rec}_2 
\end{align}
\end{subequations}
Note that the system of equations (\ref{eq:ei_rec1b}) has the a similar form to (\ref{eq:ei_ionization_ener}).

\section{Collisional-radiative modeling using the multifluid equations}
\label{sec:CR}
Before presenting the numerical results, we briefly describe how to apply the previous formulation of the rates to construct CR models in the context of the multifluid equations. We first note that the same set of atomic data and cross sections is required as in standard CR model. The only difference is that the rates now include corrections due to the multifluid effect. Hence for a given set of data, the results obtained using the multifluid model will approach the standard (single-fluid) results in the limit of $\lambda \rightarrow 0$. This can be seen easily from the fact that all the expressions of the multifluid rates converge to single-fluid results in same limit. We will also demonstrate this convergence in sec. \ref{sec:numerics} via numerical calculations.

Let us now consider an example of an atomic hydrogen plasma, which contains H, H$^{+}$ and the free electrons $e$. The neutral atom H can have many bound states, the interactions between which can occur via a number of processes. In addition, ionization can proceed from those atomic levels by collisions with the free electrons. Consider now a three-fluid model (neutral-ion-electron) where the all atomic states of H belong to the same fluid (neutral). We also assume that the VDF of each fluid is a perfect Maxwellian so transport fluxes can be omitted. In this case, we end up with three sets of fluid equations (Euler), one for each fluid. \cite{burgers_flow_1969} For neutral H, the fluid equations must be extended to the multi-species Euler equations to accommodate different atomic states of H. The excitation/deexcitation rates between these atomic states can be constructed following our previous work\cite{le_modeling_2015}. The ionization/recombination rate for each atomic level can be computed using the formulas in sections \ref{sec:ion} and \ref{sec:rec}. For example, with 10 atomic levels, one would need to compute the rates for 90 excitation/deexcitation transitions and 20 ionization/recombination transitions. All these rates are tabulated as functions of $\lambda$ and $T_e$. During the calculation, the rates for a specific condition can be obtained by interpolation. In addition, one can also compute momentum and energy exchange rate coefficients in a similar fashion. Although we have only discussed electron induced excitation and ionization processes, other processes can also be incorporated in a consistent manner.

Generalization to multiply charged ions is also straight forward. Let us consider an example of Helium where the plasma contains He, He$^{+}$, He$^{++}$ and $e$. In the simplest three-fluid formulation, we can treat He as a neutral fluid, He$^{+}$ and He$^{++}$ together as an ion fluid, and the free electrons as an electron fluid. Since He and He$^{+}$ also include multiple excited states, the neutral and the ion fluid equations are extended to multi-species Euler equations. The exchange rates (number densities, momentum and energy) for excitation and ionization (and their reverses) can be constructed similarly. In all cases, we also need to consider elastic collisions between different fluids: electron-ion, electron-neutral, ion-neutral. These will appear through the momentum and energy equations of all the fluids. Note that here the collision between He$^+$ and He$^{++}$ are omitted because they belong to the same fluid. In the case where each charge state is considered as an individual fluid, we end up with a four-fluid model, and He$^+$-He$^{++}$ collision now must be taken into account. In the presence of hot electrons, we can treat the bulk and the hot electrons as two separate fluids in a straight-forward manner.

\section{Numerical results}\label{sec:numerics}
\subsection{Exchange rates}
In this section, the numerical results of the reaction rates are presented. We consider a partially ionized hydrogen plasma with neutrals, ions and free electrons. The neutral atomic states are defined according to the principle quantum number $n$ and the energy levels are given from the Bohr model, e.g. $E_n = I_H (1-1/n^2)$ where $I_H = 13.6$ eV is the ionization energy of the ground state. The differential ionization cross section of state $n$ is defined according to the semi-classical model\cite{zeldovich_physics_2002}:
\begin{align}
\frac{d \sigma^{ion}_n}{d\Upsilon} = \frac{4 \pi a_0^2 I_H^2}{\Upsilon^2} \frac{1}{\varepsilon} \quad \textrm{s.t.} \quad \bar{\sigma}^{ion}_n = \left( 4 \pi a_0^2 \right) \frac{I_H^2 \left( \varepsilon - I_n \right)}{I_n \varepsilon^2}
\end{align}
where $I_n = I_H-E_n$ and $a_0 = 0.529$ $\textrm{\AA}$ is the Bohr radius. 

\begin{figure}
\centering
\includegraphics[scale=.8]{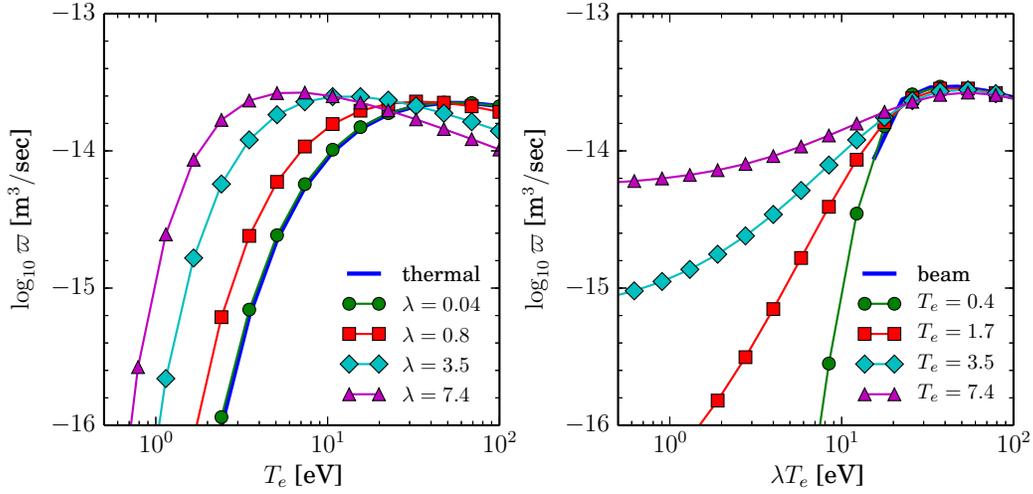}
\caption{Multifluid reaction rates for electron induced ionization collision. The solid lines correspond to the two asymptotic limits: thermal $(\lambda \rightarrow 0)$ and beam ($T_e \rightarrow 0$).}
\label{fig:ion0_0}
\end{figure}

\begin{figure}
\centering
\includegraphics[scale=.8]{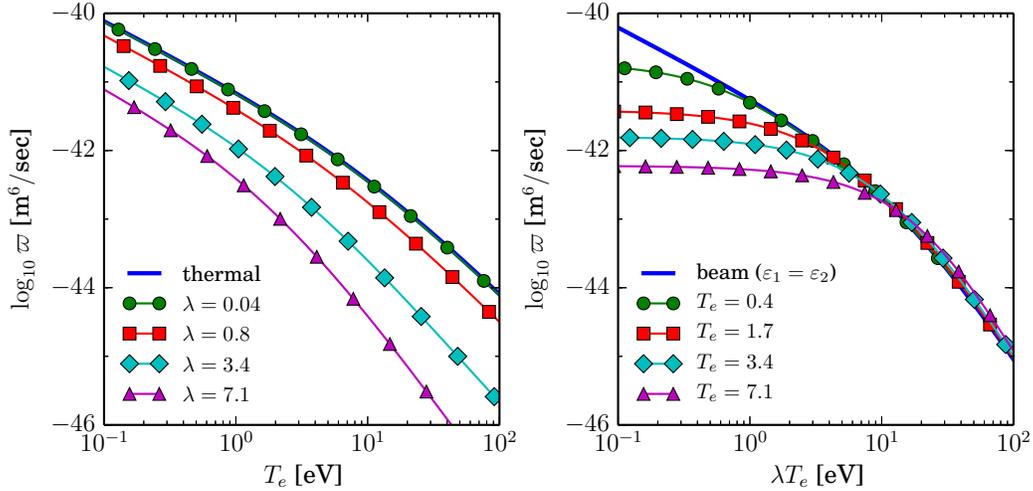}
\caption{Multifluid reaction rates for electron induced recombination collision. The solid lines correspond to the two asymptotic limits: thermal $(\lambda \rightarrow 0)$ and beam ($T_e \rightarrow 0$). The beam limit is computed for $\varepsilon_1 = \varepsilon_2$, i.e., the scattered and ejected electrons share equal amount of energy.}
\label{fig:rec0_0}
\end{figure}

The numerical integrations of all the exchange rate coefficients are carried out using the adaptive algorithm from the cubature package \cite{johnson_cubature_2014}. These results are also compared with Monte Carlo integrations of the full transfer integral with excellent agreement. For brevity, we only show the results for zeroth-order reaction rates. Figures \ref{fig:ion0_0} and \ref{fig:rec0_0} show the ionization and recombination rates of ground state hydrogen for an electron induced collision with different values of $\lambda$. It must be noted that $\lambda$ refers to the relative drift between H and $e$ for ionization, and H$^+$ and $e$ for recombination. For simplicity, H and H$^+$ are treated as the same fluid in our next calculations, so $\lambda$ is the same for both processes. The results from figures \ref{fig:ion0_0} and \ref{fig:rec0_0} confirm that both thermal (single-fluid) and beam asymptotic limits of the rates are recovered from the derived expressions. Figure \ref{fig:ion0_0} also indicates that the relative drift between two fluids (measured by $\lambda$) can increase the ionization rates at low temperature; this observation is similar to the case of excitation/deexcitation. On the contrary, figure \ref{fig:rec0_0} suggests that the recombination rates get weaker as $\lambda$ increases. It must be noted that the standard Saha relation (macroscopic) is only satisfied in the thermal limit. For $\lambda \neq 0$, detailed balance is enforced through the Fowler relation (microscopic).

\subsection{Collisional-radiative rate equations}
The multifluid reaction rates from the previous section are used to solve the collisional-radiative (CR) rate equations. In the first test, we consider an isothermal system of atomic hydrogen plasma with constant electron number density. A total number of 10 atomic states of H is used in the calculation in addition to H$^+$. The parameter $\lambda$ is introduced as a constant to examine the multifluid effect. This relative drift can be realized in a system where there is a steady state current. For example, in the magnetohydrodynamic (MHD) limit \cite{rosa_magnetohydrodynamic_1968}, the plasma current $\bvec{J}$ can be approximated by $ \bvec{J}  \simeq \frac{1}{\eta} \left( \bvec{E} + \bvec{u} \times \bvec{B} \right)$ where $\eta$ is the plasma resistivity. To make the problem more realistic, we also include line radiation between bound states and further assume that the plasma is optically thin. 

\begin{figure}
\centering
\includegraphics[scale=.8]{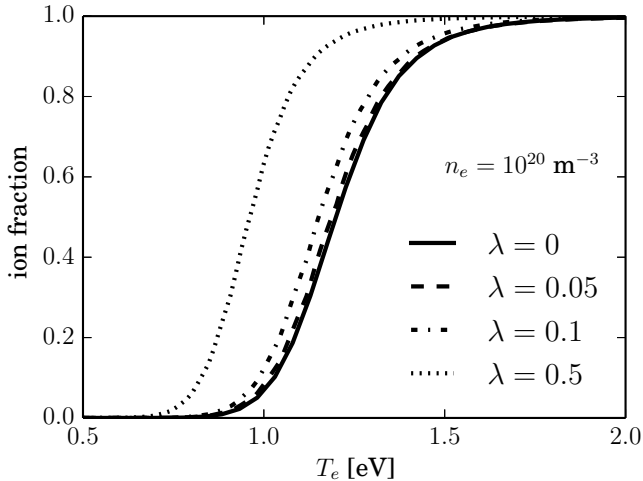}
\caption{Ion fraction vs $T_e$ for the case of atomic hydrogen plasma with $n_e = 10^{20}$ m$^{-3}$ and different values of the multifluid $\lambda$ parameters. The solutions are obtained by solving the steady-state rate equations for fixed values of $n_e$, $T_e$ and $\lambda$. Line radiation is included and the plasma is assumed to be optically thin.}
\label{fig:ion_frac}
\end{figure}

The resultant system of rate equations can be put into the following form:
\begin{align}\label{eq:cr}
\frac{d \bvec{\tilde{n}}}{dt} = \overline{\bvec{R}} \cdot \bvec{\tilde{n}}
\end{align}
where $\bvec{\tilde{n}}$ is the state population vector and $\overline{\bvec{R}}$ is the rate matrix. For constant $n_e$, $T_e$ and $\lambda$, $\overline{\bvec{R}}$ is also constant. The steady-state solutions of (\ref{eq:cr}) can be obtained by setting $\frac{d \bvec{\tilde{n}}}{dt} = 0$, and solving $\overline{\bvec{R}} \cdot \bvec{\tilde{n}} = 0$. In order to avoid the trivial solution of $\bvec{\tilde{n}} = 0$, charge neutrality is used as a constraint. Equation (\ref{eq:cr}) is solved for a range of $(n_e,T_e,\lambda)$. Figure \ref{fig:ion_frac} shows the resultant ion fraction for the case of $n_e = 10^{20}$ m$^{-3}$. It can be seen that the ion fraction deviates from the single-fluid result when $\lambda \neq 0$. We note that the solutions plotted in figure \ref{fig:ion_frac} are different from the LTE solutions since line radiation is included in the system. Furthermore, when $\lambda \neq 0$, the forward and the backward rates of the inelastic processes also deviate from the standard Boltzmann/Saha relation.

In the next test, we consider an isochoric system of a two-fluid hydrogen plasma (electrons and heavy particles). Since the system is closed, the momentum densities and temperatures of the two fluids are coupled to the rate equations for number densities and evolved self-consistently. We assume that the all the heavy particles (neutrals and ions) belong to the same fluid, so that the momentum and energy exchange processes between these particles are infinitely fast. The governing equations for this system are the same as the ones described in our previous paper\cite{le_modeling_2015} (see appendix D) but with additional terms due to ionization/recombination. The initial conditions of these simulations are listed in table \ref{tab:IC}. Initially, all the atoms are at rest, and the atomic states are in Boltzmann equilibrium at 0.3 eV. A fraction of hot electrons at $T_e = 3$ eV is added, and their mean velocities are varied to demonstrate the multifluid effect. The ion density follows from charge neutrality.

\begin{table}
\begin{tabular}{|c|c|c|}
\hline \hline
 & number density & temperature \\ 
\hline 
atomic states & $n_k = 0.9 \mathcal{B}_k n_t$ for $k=1-10$& 0.3 eV\\ 
 \hline
  ion    & $n_i = 0.1n_t$ & 0.3 eV\\ 
\hline
electron & $n_e = 0.1n_t$ & 3 eV\\ 
\hline \hline
\end{tabular} 
\caption{Initial conditions of 0D test cases. The total atomic density $n_t$ is $10^{20}$ m$^{-3}$. The atomic states are initialized according to a Boltzmann distribution at $T_h$, i.e., $\mathcal{B}_k = \frac{\mathpzc{g}_k e^{-E_k/T_h}}{\mathbb{Z}_n}$ where $\mathbb{Z}_n$ is the electronic partition function.}
\label{tab:IC}
\end{table}

\begin{figure}
\centering
\includegraphics[scale=.9]{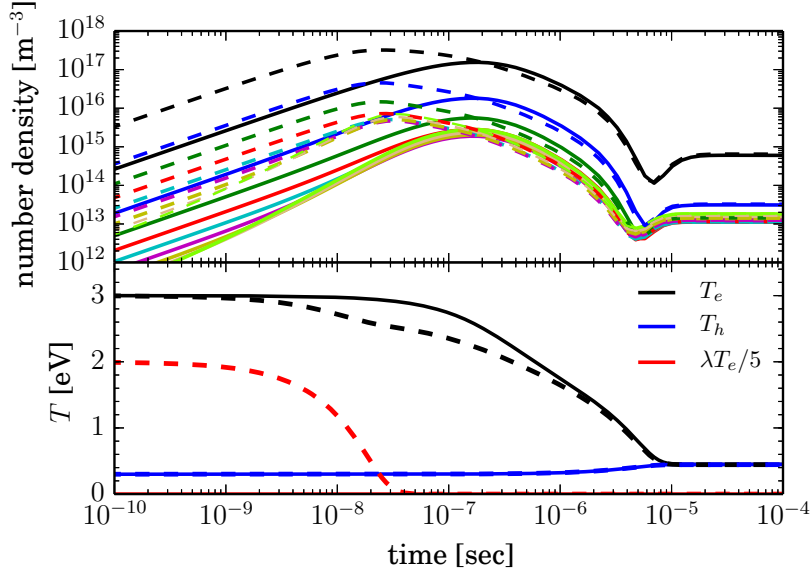}
\caption{Time evolution of number densities of the atomic states ($2-10$) and temperatures for the two test cases with initial conditions from table \ref{tab:IC}. Solid lines denote case I with $\lambda = 0.01$ initially and dashed lines denote case II with $\lambda = 3.3$ initially. The evolution of the equivalent drift temperature $\lambda T_e$ (red) is shown for case II. For case I, $\lambda T_e \simeq 0$, which corresponds to a single-fluid calculation.}
\label{fig:NTevol_56}
\end{figure}

Figure \ref{fig:NTevol_56} shows the time evolution of the atomic state density (top) and the temperatures (bottom) of two different cases. Case I, shown in solid lines, corresponds to an initial zero relative drift velocity ($\lambda = 0$) and case II, shown in dashed lines, to a large initial relative drift velocity ($\lambda = 3.3$). Similar to the observation made when considering excitation/deexcitation only, the kinetics of inelastic collisions is enhanced when the relative drift between the two fluids is significant. This is indicated by an early increase in the population of the excitation states from figure \ref{fig:NTevol_56}. Moreover, the temperature relaxation between two cases are also different as can be seen from the bottom plot of figure \ref{fig:NTevol_56}. We remark that in this test case, the enhancement to the kinetics due to the relative drift only persists on the momentum relaxation time scale. 

\begin{figure}
\centering
\includegraphics[scale=.8]{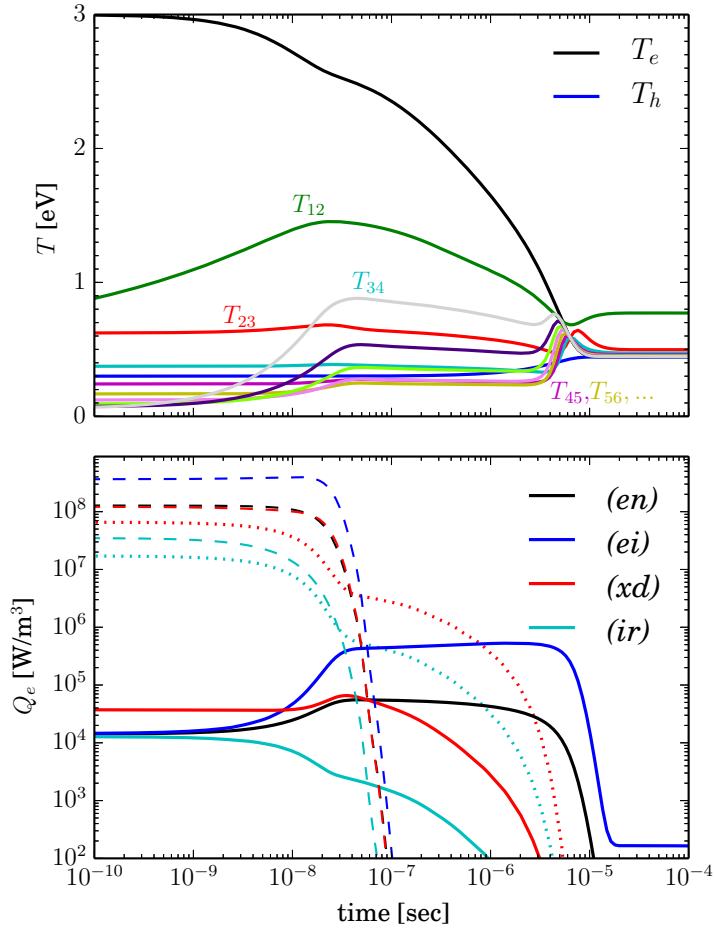}
\caption{Boltzmann temperatures of the excited states and energy exchange rates of the electrons. The Boltzmann temperatures are defined according to eq. (\ref{eq:Tb}). In the bottom plot, different colors indicate different processes: \textit{(en)} refers to electron-neutral, \textit{(ie)} to Coulomb, \textit{(xd)} to excitation/deexcitation, and \textit{(ir)} to ionization/recombination collisions. The line types (solid, dashed, dotted) are used to distinguish between different terms in the energy exchange.}
\label{fig:RT6}
\end{figure}

To further examine the relaxation process in the presence of the multifluid effect, figure \ref{fig:RT6} shows the time evolution of the Boltzmann temperatures of the excited states and the energy exchange rates due to different types of collision for case II. The Boltzmann temperatures, defined between two adjacent states $\ell$ and $u$ ($\ell < u$), are as follows:
\begin{align}\label{eq:Tb}
T_{\ell u} = \frac{E_u - E_\ell}{\ln \left( \frac{n_\ell / \mathpzc{g}_\ell}{n_u / \mathpzc{g}_u} \right)}
\end{align}
where $n_\ell$, $n_u$ are the number densities of levels $\ell$, $u$. These temperatures are used to measure deviation from the Boltzmann equilibrium of the atomic states. It can be seen from the top of figure \ref{fig:RT6} that at approximately $4 \times 10^{-6}$ sec, all the higher states ($n>3$) have reached equilibrium with the free electrons. Due to the large energy gaps between the first 3 atomic states, these states take a longer time to equilibrate, e.g., $T_{23} \simeq T_e$ at approximately $2 \times 10^{-5}$ sec. This condition is known as partial local thermodynamic equilibrium \cite{van_der_mullen_excitation_1990}. Although not shown in here, the system eventually achieves complete thermodynamic equilibrium at a much later time. 

The energy exchange rates of the electrons are shown in the bottom plot of figure \ref{fig:RT6}. The solid lines denote thermal relaxation (terms proportional to $\mathcal{J}$), the dashed lines denote frictional work (terms proportional to $\mathcal{R}$), and the dotted lines denote heat of formation due to inelastic collisions (terms proportional to $\Gamma$). In general, these terms can have different signs where positive and negative mean heating and cooling respectively. For this particular case, the electrons are losing energy due excitation/ionization and thermal relaxation with the heavy particles; therefore, the solid and the dotted lines indicate cooling rates. On the other hand, the friction between the electrons and heavy particles can do work to heat the electrons, so the dashed lines here refer to heating rates. One can see from the bottom plot of figure \ref{fig:RT6} that up to $10^{-7}$ sec, frictional heating and heat of formation are the two main energy transfer mechanisms. This also corresponds to the momentum relaxation time scale, after which the momentum of the electrons have been completely absorbed by the heavy particles, signalling a change to single-fluid kinetics. One can also note that during $10^{-8} < t < 10^{-7}$, there are competing effects between all the processes, and inelastic collisions in general can also contribute to total energy exchange and should not be neglected. Although this test case suggests that the multifluid effect only persists on the momentum relaxation time scale, we expect that this effect becomes more significant for system where there exists a steady state current (since the the drift is always maintained due to the current). This will be examined in a future publication where spatial inhomogeneity will also be included.

\section{Conclusion}\label{sec:conclusion}
We have presented a model for ionization and recombination collisions in a multifluid plasma. The model is rigorously derived from kinetic theory and follows directly from our previous work on the modeling of excitation and deexcitation collisions\cite{le_modeling_2015}. The derived exchange coefficients are shown to have proper asymptotic limits, and satisfy the principle of detailed balance. Using the new set of rate coefficients, we have developed and tested a new multifluid collisional-radiative model for atomic hydrogen with semi-classical cross sections for all the elementary processes. This model has two important features: (a) multifluid effect is captured in the definitions of the rate, and (b) the momentum and energy exchanges due to inelastic collisions are included.

Numerical calculations of the exchange rates are carried out and the accuracy is confirmed with direct Monte Carlo integration. The results indicate that in the presence of a relative drift between two reactant fluids, the rates can be significantly different than the single-fluid limit. Two numerical tests are conducted demonstrate the capability of the new model. In the first test, we compare the steady-state solutions of the collisional-radiative rate equations with constant $n_e$, $T_e$ and $\lambda$. The results converge to the single-fluid solution as $\lambda \rightarrow 0$, and can deviate from that when $\lambda \neq 0$. In the second test, an isochoric heating of a partially ionized hydrogen plasma is performed in a virtual test cell to demonstrate the coupling between various collision processes. We observe that in general inelastic collisions can participate in the overall energy exchange process and should be included in the model. The present work can be extended to other types of collision, e.g., charge exchange and molecular collisions, with slight modifications. Future work focuses on examining the nonlinear coupling of transport with collisional-radiative kinetics by means of the multifluid transport equations \cite{braginskii_transport_1965}.  

\section*{Acknowledgments}
Research was supported by the Air Force Office of Scientific Research (AFOSR), grant numbers 14RQ05COR (PM: Dr. J. Marshall) and 14RQ13COR (PM: M. Birkan).

\numberwithin{equation}{section}
\begin{appendices}
\section{Separation of variables}
\label{app:threebody} 
Similarly to excitation, the ionization process has two particles in the initial istate, but the final state includes a third particle, since an electron extracted from the target to yield an ion state ($t\rightarrow i + e$). The process is therefore:
\begin{align}\label{eq:e1}
s(\bvec{v}_{s_0}) + t(\bvec{v}_{t_0}) \Leftrightarrow s(\bvec{v}_{s_1}) + i(\bvec{v}_{i_2}) + e(\bvec{v}_{e_2})
\end{align} 
In the case of ionization, one must integrate over the distribution functions of the initial variables, which remain $s,t$, and the procedure used in an excitation collision remains valid. However, for recombination, we have a triple product of VDFs:
\begin{align}\label{eq:e2}
f_s(\bvec{v}_{s_1})\,f_i(\bvec{v}_{i_2})\,f_e(\bvec{v}_{e_2}) = &\left(\frac{m_s}{2\pi T_s}\right)^{\frac{3}{2}} \left(\frac{m_i}{2\pi T_i}\right)^{\frac{3}{2}} \left(\frac{m_e}{2\pi T_e}\right)^{\frac{3}{2}} \exp \left[ \mathcal{A} \right]
\end{align}
The argument of the exponential function is:
\begin{align}\label{eq:Aeis}
\mathcal{A} =\beta_s (\bvec{v}_{s_1}\!-\!\bvec{u}_s)^2 + \underbrace{ \beta_e (\bvec{v}_{e_2}\!-\!\bvec{u}_e)^2 + \beta_i (\bvec{v}_{i_2}\!-\!\bvec{u}_i)^2 }_{\mathcal{A}_{ei}}
\end{align}
where $\beta_s = \frac{m_s}{2T_s}$. In order to perform the separation of variables, it is necessary to proceed in two steps. Thus, we can consider the ionization process as follows:
\begin{enumerate}[label=\emph{\alph*})]
\item the formation of an excited state $t^*$ via scattering: $s(\bvec{v}_{s_0}) + t(\bvec{v}_{t_0}) \Rightarrow s (\bvec{v}_{s_1}) + t^*(\bvec{v}_{t_1})$
\item the spontaneous ionization of the $t^*$ state into ion and electron: $t^*(\bvec{v}_{t_1}) \Rightarrow e(\bvec{v}_{e_2}) + i(\bvec{v}_{i_2})$
\end{enumerate}
The reverse process, recombination, would similarly follow two steps:
\begin{enumerate}[label=\emph{\alph*})]
\item the formation of an excited state $t^*$ via recombination: $e(\bvec{v}_{e_2}) + i(\bvec{v}_{i_2}) \Rightarrow  t^* (\bvec{v}_{t_1})$ 
\item the spontaneous deexcitation of the $t^*$ state via scattering: $s(\bvec{v}_{s_1}) + t^*(\bvec{v}_{t_1}) \Rightarrow s(\bvec{v}_s) + t(\bvec{v}_t)$
\end{enumerate}
Consider now the first part of this two-step recombination process, which involves the product of the two VDFs for electron and ion: $f_e(\bvec{v}_{e_2}) \cdot f_i(\bvec{v}_{i_2})$.
The argument of the exponential function resulting from this product is $\mathcal{A}_{ei}$ as defined in (\ref{eq:Aeis}). Let us first perform the separation of variables for the product $f_e \cdot f_i$ (see Appendix B of Le \& Cambier \cite{le_modeling_2015}), such that the argument becomes:
\begin{equation}\label{eq:e12}
\mathcal{A}_{ei} = (\beta_e\!+\!\beta_i)\bvec{C_t}^2 + \frac{\beta_e\beta_i}{\beta_e\!+\!\beta_i} \tvec{g}_2^2
\end{equation}
where 
\begin{subequations}
\begin{align}
\label{eq:Ct}
\bvec{C}_t &= \bvec{v}_{t_1} \!-\!\bvec{u}_{t_1} + \gamma_t \tvec{g}_2\\
\tvec{g}_2 &= \bvec{g}_2 - \bvec{w}_2 \\
\bvec{v}_{t_1} &= \frac{m_e\bvec{v}_{e_2}\!+\!m_i\bvec{v}_{i_2}}{m_t}\\
\bvec{u}_{t_1} &= \frac{m_e\bvec{u}_{e}\!+\!m_i\bvec{u}_{i}}{m_t}\\
\gamma_t &= \frac{1}{\beta_e + \beta_i}\left(\beta_e\frac{m_i}{m_t}-\beta_i\frac{m_e}{m_t}\right)
\end{align}
\end{subequations}
and the relative velocity $\bvec{g}_2$ is defined according to (\ref{eq:transformation}).

We can now multiply by the VDF for the scattering particle for the second step of the recombination process. This leads to the total argument:
\begin{equation}\label{eq:e13}
\mathcal{A} = (\beta_e\!+\!\beta_i)\bvec{C}_t^2 + \frac{\beta_e\beta_i}{\beta_e\!+\!\beta_i} \tvec{g}_2^2 + \beta_s (\bvec{v}_{s_1}\!-\!\bvec{u}_s)^2
\end{equation}
Let us also define
\begin{equation}\label{eq:e15}
\bvec{V}^* = \bvec{v}_{t_1} - \bvec{u}_{t_1} = \bvec{V}-\bvec{U}_1-\frac{m_s}{M}\tvec{g}_1
\end{equation}
with $\tvec{g}_1 = \bvec{g}_1\!-\!\bvec{w}_1$. This yields:
\begin{equation}\label{eq:e16}
\bvec{v}_{s_1} -\bvec{u}_s = \bvec{V}\!-\!\bvec{U}_1+\frac{m_t}{M}\tvec{g}_1= \bvec{V}^*+\tvec{g}_1
\end{equation}
and, from (\ref{eq:Ct}), 
\begin{equation}\label{eq:e17}
\bvec{C}_t =  \bvec{V}^*+\gamma_t \,\tvec{g}_2
\end{equation}
Inserting into (\ref{eq:e13}):
\begin{align}\label{eq:e18}
\mathcal{A} = (\beta_s\!+\!\beta_e\!+\!\beta_i)\bvec{V}^{*2} & + \beta_s \tvec{g}_1^2\nonumber\\
             & +\left[(\beta_e\!+\!\beta_i)\gamma_t^2 + \frac{\beta_e\beta_i}{\beta_e\!+\!\beta_i}\right] \tvec{g}_2^2\\
             & + 2\gamma_t(\beta_e\!+\beta_i)\bvec{V}^*\cdot \tvec{g}_2  +2\beta_s \bvec{V}^*\cdot \tvec{g}_1 \nonumber
\end{align}
Let us now try the following variable substitution
\begin{equation}\label{eq:e19}
\bvec{V}^{**} = \bvec{V}^* +\tilde{\gamma} \tvec{g}_2 + \tilde{\delta} \tvec{g}_1
\end{equation}
Thus,
\begin{align*}
\bvec{V}^{**2} &= \bvec{V}^{*2} +\tilde{\gamma}^2 \tvec{g}_2^2 +\tilde{\delta}^2 \tvec{g}_1^2 \\
 &+2\tilde{\gamma}\bvec{V}^*\cdot \tvec{g}_2  + 2\tilde{\delta}\bvec{V}^*\cdot \tvec{g}_1 +2\tilde{\gamma}\tilde{\delta} \tvec{g}_1\cdot \tvec{g}_2
\end{align*}
Defining $\Sigma_\beta=\beta_s\!+\!\beta_s\!+\!\beta_i$ and choosing
\begin{equation}\label{eq:e20}
\tilde{\delta} = \frac{\beta_s}{\Sigma_\beta}, \qquad \tilde{\gamma}=\frac{\beta_e\!+\!\beta_i}{\Sigma_\beta}\gamma_t
\end{equation}
we obtain
\begin{align*}
\Sigma_\beta\bvec{V}^{**2} = &\Sigma_\beta\bvec{V}^{*2} 
  + \frac{\beta_s^2}{\Sigma_\beta}  \tvec{g}_1^2 +\frac{(\beta_e\!+\!\beta_i)^2}{\Sigma_\beta}\gamma_t^2 \tvec{g}_2^2\\
&+2\gamma_t(\beta_e\!+\!\beta_i)\bvec{V}^*\cdot \tvec{g}_2 +2\beta_s\bvec{V}^*\cdot \tvec{g}_1
+ 2\gamma_t\frac{\beta_s(\beta_e\!+\!\beta_i)}{\Sigma_\beta} \tvec{g}_1\cdot \tvec{g}_2
\end{align*}
Comparing with (\ref{eq:e18}), we can simplify the argument as:
\begin{align}\label{eq:e21}
\mathcal{A} =&\Sigma_\beta\bvec{V}^{**2} + \left[ \frac{\beta_s(\beta_e\!+\!\beta_i)}{\Sigma_\beta}\gamma_t^2 + \frac{\beta_e\beta_i}{\beta_e\!+\!\beta_i} \right] \tvec{g}_2^2\\
 & +\frac{\beta_s(\beta_e\!+\!\beta_i)}{\Sigma_\beta} ]\left[ \tvec{g}_1^2 -2\gamma_t \tvec{g}_1 \cdot \tvec{g}_2 \right]\nonumber
\end{align}
Define now
\begin{equation}\label{eq:e22}
\bvec{j} = \tvec{g}_1 - \gamma_t \tvec{g}_2
\end{equation}
We can now eliminate the last dot product, since $\tvec{g}_1^2 \!-\! 2\gamma_t \tvec{g}_1 \cdot \tvec{g}_2 \! \!= \bvec{j}^2 \!-\!\gamma_t^2 \tvec{g}_2^2$.
Inserting into (\ref{eq:e21}), we finally obtain:
\begin{equation}\label{eq:e23}
\mathcal{A} =(\beta_s\!+\!\beta_e\!+\!\beta_i)\bvec{V}^{**2} + \frac{\beta_e\beta_i}{\beta_e\!+\!\beta_i} \tvec{g}_2^2 + \frac{\beta_s(\beta_e\!+\!\beta_i)}{\beta_s\!+\!\beta_e\!+\!\beta_i} \bvec{j}^2
\end{equation}
Here all dot products have been removed with the proper change of variables. One can also show that:
\begin{align}\label{eq:e24}
\beta_s+\beta_e+\beta_i &= \frac{M}{2} \frac{m_s T_e T_i + m_e T_s T_i + m_i T_s T_e}{MT_s T_e T_i} \equiv \frac{M}{2T^*}\\
\frac{\beta_e \beta_i}{\beta_e + \beta_i} &= \frac{m_e m_i}{2(m_e + m_i)} \frac{m_e+m_i}{m_e T_i + m_i T_e} \equiv \frac{\mu_t}{2 \tilde{T}_t}\\
\frac{\beta_s (\beta_e + \beta_i)}{\beta_s+\beta_e+\beta_i} &= \frac{m_s (m_e +m_i)}{2M} \frac{M\tilde{T}_t}{m_s T_e T_i + m_e T_s T_i + m_i T_s T_e}  \equiv \frac{\mu}{2 \tilde{T}}
\end{align}
where
\begin{align}\label{eq:e25}
T^* &= \frac{MT_s T_e T_i}{m_s T_e T_i + m_e T_s T_i + m_i T_s T_e} \\ 
\tilde{T}_t & = \frac{m_e T_i + m_i T_e}{m_e+m_i}\\
\tilde{T} &= \frac{m_s T_e T_i + m_e T_s T_i + m_i T_s T_e} {M\tilde{T}_t} \\
\mu_t & = \frac{m_e m_i}{m_e + m_i} \\
\mu & = \frac{m_s (m_e +m_i)}{M}
\end{align}
The product of the three Maxwellian VDF becomes:
\begin{equation}\label{eq:e26}
\begin{split}
f_s (\mathbf{v}_{s_1}) \cdot f_e (\mathbf{v}_{e_2}) \cdot f_i (\mathbf{v}_{i_2}) = 
\left(\frac{M}{2\pi T^*}\right)^{\frac{3}{2}} \!\exp\left[-\frac{M\bvec{V}^{**2}}{2T^*}\right] \cdot  
\left(\frac{\mu_t}{2\pi \tilde{T}_t}\right)^{\frac{3}{2}} \!\exp\left[-\frac{\mu_t\tvec{g}_2^{2}}{2\tilde{T}_t}\right] \cdot 
\\ \left(\frac{\mu}{2\pi \tilde{T}}\right)^{\frac{3}{2}} \!\exp\left[-\frac{\mu\bvec{j}^2}{2\tilde{T}}\right] \equiv f^{**}(\bvec{V^{**}})\cdot \tilde{f}_t (\tvec{g}_2) \cdot \tilde{f} (\bvec{j})
\end{split}
\end{equation}
All subsequent expressions can now be simplified with this separation of variables. For example, any operator $\mathscr{O}$ that depends only on variables expressed using the relative velocities ($\bvec{g}_0,\bvec{g}_1,\bvec{g}_2$), we have:
\begin{align}\label{eq:b17}
\int d^3\bvec{v}_{s_1} d^3\bvec{v}_{e_2} d^3\bvec{v}_{i_2} f_s f_e f_i \mathscr{O}(\bvec{g}_0,\bvec{g}_1,\bvec{g}_2) 
= \underbrace{\int d^3\bvec{V}^{**} f^{**}(\bvec{V}^{**})}_{\equiv 1} \cdot \int d^3 \tvec{g}_1 d^3\tvec{g}_2 \tilde{f}_t (\tvec{g}_2) \tilde{f} (\bvec{j}) \mathscr{O}(\bvec{g}_0,\bvec{g}_1,\bvec{g}_2)
\end{align}

\section{Exchange coefficients for ionization}
\label{app:ion} 
We describe in this appendix various exchange terms computed from the transfer integral for an ionization collision, starting from the transfer integral given in eq. (\ref{eq:psi_ion_st}). The exchange variables during the collision can be expressed in terms of $\bvec{V}^*$, $\bvec{g}_0$, $\bvec{g}_1$ and $\bvec{g}_2$. Since $f_{V^*}$ represents a Maxwellian VDF centered at zero, we have $\int d^3 \bvec{V}^* f_{V^*} = 1$, $\int d^3 \bvec{V}^* \, \bvec{V}^* \, f_{V^*} = 0$ and $\int d^3\bvec{V}^* \, \frac{1}{2} M \bvec{V}^{*2} \, f_{V^*} = \frac{3}{2} T^*$. Thus, if $\psi$ is independent of $\bvec{V}^*$, we can eliminate the integral over $\bvec{V}^*$. For the case where $\psi$ is linear in $\bvec{V}^*$, the transfer integral goes to zero. Here the subscripts $st$ in the differential cross sections are omitted for brevity.

Let us now consider the case where $\psi = \bvec{g}_p $ ($p=0,1,2)$. As shown in Le \& Cambier\cite{le_modeling_2015}, the only non-zero velocity component survived after the integration is the one parallel to the relative drift velocity $\bvec{w}_0$. Using the definition from (\ref{eq:ion_R}), the friction coefficients can be written as follows:
\begin{subequations}\label{eq:friction_coefs}
\begin{align}
R^{ion}_{0} &= \frac{2}{3} n_s n_t \bar{g}_{\tilde{T}} e^{-\lambda} \int_{x^*}^\infty \! dx_0 \, x_0^{2} \, e^{-x_0}\, \zeta^{(1)} \left( \sqrt{\lambda x_0} \right) \, \overline{\sigma}^{ion}\\
R^{ion}_{1} &= \frac{2}{3} n_s n_t \bar{g}_{\tilde{T}} e^{-\lambda} \int_{x^*}^\infty \! dx_0 \,x_0^{\frac{3}{2}}\,e^{-x_0}\, \zeta^{(1)} \left( \sqrt{\lambda x_0} \right) \, \int_{x^*}^{x_0}  \sqrt{x_1} \avom{ c_{\chi_1}} \frac{d {\sigma}^{ion}}{d\upsilon} d\upsilon\\
R^{ion}_{2} &= \frac{2}{3} n_s n_t \bar{g}_{\tilde{T}} e^{-\lambda} \int_{x^*}^\infty \! dx_0 \,x_0^{\frac{3}{2}}\,e^{-x_0}\, \zeta^{(1)} \left( \sqrt{\lambda x_0} \right) \, \int_{x^*}^{x_0}  \sqrt{x_2} \avom{  c_{\chi_2}} \frac{d {\sigma}^{ion}}{d\upsilon} d\upsilon
\end{align}
\end{subequations}
where $\zeta^{(1)} (\xi) = \frac{3}{4\xi^2} \left[ \cosh (2\xi) - \frac{\sinh (2\xi)}{2\xi} \right]$ and $\lim_{\xi \rightarrow 0} \zeta^{(1)} (\xi) = 1$. For isotropic scattering, i.e., $\mathcal{G}^{ion} =$ constant, $R^{ion}_{1}=R^{ion}_{2}=0$. 

For the case of $\psi = \bvec{g}_p \cdot \bvec{g}_q $ ($p,q=0,1,2)$, we arrive at the following thermal relaxation coefficients using the definitions in (\ref{eq:ion_J}):
\begin{subequations}\label{eq:thermal_coefs}
\begin{align}
J^{ion}_{00} &=   n_s n_t \bar{g}_{\tilde{T}} e^{-\lambda} \int_{x^*}^\infty \!dx_0\,x_0^2  \,e^{-x_0}\,\zeta^{(0)} \left( \sqrt{\lambda x_0} \right) \,  \bar{\sigma}^{ion} \\
J^{ion}_{11} &=   n_s n_t \bar{g}_{\tilde{T}} e^{-\lambda} \int_{x^*}^\infty \!dx_0\,x_0  \,e^{-x_0}\,\zeta^{(0)} \left( \sqrt{\lambda x_0} \right)  \int_{x^*}^{x_0}  x_1  \frac{d{\sigma}^{ion}}{d\upsilon}  d\upsilon\\
J^{ion}_{22} &=   n_s n_t \bar{g}_{\tilde{T}} e^{-\lambda} \int_{x^*}^\infty \!dx_0\,x_0  \,e^{-x_0}\,\zeta^{(0)} \left( \sqrt{\lambda x_0} \right) \int_{x^*}^{x_0}  x_2 \frac{d{\sigma}^{ion}}{d\upsilon}   d\upsilon\\
J^{ion}_{01} &=   n_s n_t  \bar{g}_{\tilde{T}} e^{-\lambda} \int_{x^*}^\infty \!dx_0\,x_0  \,e^{-x_0}\,\zeta^{(0)} \left( \sqrt{\lambda x_0} \right) \int_{x^*}^{x_0}  \sqrt{x_0 x_1} \avom{ c_{\chi_1} } \frac{d{\sigma}^{ion}}{d\upsilon}  d\upsilon\\
J^{ion}_{02} &=  n_s n_t \bar{g}_{\tilde{T}} e^{-\lambda} \int_{x^*}^\infty \!dx_0\,x_0  \,e^{-x_0}\,\zeta^{(0)} \left( \sqrt{\lambda x_0} \right)  \int_{x^*}^{x_0}   \sqrt{x_0 x_2} \avom{c_{\chi_2}} \frac{d{\sigma}^{ion}}{d\upsilon} d\upsilon\\
J^{ion}_{12} &=  n_s n_t \bar{g}_{\tilde{T}} e^{-\lambda} \int_{x^*}^\infty \!dx_0\,x_0  \,e^{-x_0}\,\zeta^{(0)} \left( \sqrt{\lambda x_0} \right) \int_{x^*}^{x_0}  \sqrt{x_1 x_2} \avom{ c_{\chi_1}c_{\chi_2} } \frac{d{\sigma}^{ion}}{d\upsilon}  d\upsilon
\end{align}
\end{subequations}
where we have used the result $\avom{s_{\chi_1} s_{\chi_2} c_{\phi_1 - \phi_2}} = 0$, since the scattering is isotropic in $\phi_1$ and $\phi_2$. Note that energy conservation implies that $x_1 = x_0 - \upsilon$ and $x_2 = \upsilon - x^*$.
 
\section{Exchange coefficients for recombination}
\label{app:rec} 
We describe in this appendix various exchange terms computed from the transfer integral for recombination processes. Here we only consider electron induced recombination with isotropic scattering. For the case of zeroth order moment ($\psi = 1$), we arrive at eq. (\ref{eq:rec_Gamma}).

Let us now consider the case where $\psi = \bvec{g}_p $ ($p=0,1,2$). It can be shown that the only non-zero velocity component survived after the integration is the one parallel to the relative drift velocity $\bvec{w}_1$. This is due to the fact that $\avom{F_2 c_{\phi_1}} = \avom{F_2 c_{\phi_2}} = 0$. Using the definition from (\ref{eq:rec_R}), the friction coefficients can be written as follows:
\begin{subequations}\label{eq:friction_coefs_rec}
\begin{align}
R^{rec}_{0} &= 0\\
R^{rec}_{1} &= \frac{2}{3} \frac{\mathpzc{g}_n}{2 \mathpzc{g}_i \mathbb{Z}_e} n_i n_e^2 \bar{g}_{\tilde{T}}  e^{-2 \lambda} e^{x^*} \int_{x^*}^\infty dx_0 \, e^{-x_0} \cdot x_0 \, \int_{x^*}^{x_0} x_1 \, \zeta^{(1)} \left( \sqrt{\lambda x_1} \right) \zeta^{(0)} \left( \sqrt{\lambda x_2} \right) \, \frac{d\sigma^{ion}}{d\upsilon} d \upsilon \\
R^{rec}_{2} &= \frac{2}{3} \frac{\mathpzc{g}_n}{2 \mathpzc{g}_i \mathbb{Z}_e} n_i n_e^2 \bar{g}_{\tilde{T}}  e^{-2 \lambda} e^{x^*} \int_{x^*}^\infty dx_0 \, e^{-x_0} \cdot x_0 \, \int_{x^*}^{x_0} x_2 \,\zeta^{(0)} \left( \sqrt{\lambda x_1} \right) \zeta^{(1)} \left( \sqrt{\lambda x_2} \right) \, \frac{d\sigma^{ion}}{d\upsilon} d \upsilon
\end{align}
\end{subequations}

For the case of $\psi = \bvec{g}_p \cdot \bvec{g}_q $ ($p,q=0,1,2)$, we arrive at the following thermal relaxation coefficients using the definition in (\ref{eq:rec_J}):
\begin{subequations}\label{eq:thermal_coefs_rec}
\begin{align}
J^{rec}_{00} &=   \frac{\mathpzc{g}_n}{2 \mathpzc{g}_i \mathbb{Z}_e} n_i n_e^2 \bar{g}_{\tilde{T}} e^{-2 \lambda} e^{x^*} \int_{x^*}^\infty dx_0 \, e^{-x_0} \cdot x_0^2 \, \int_{x^*}^{x_0}  \zeta^{(0)} \left( \sqrt{\lambda x_1} \right) \zeta^{(0)} \left( \sqrt{\lambda x_2} \right) \, \frac{d\sigma^{ion}}{d\upsilon} d \upsilon \\
J^{rec}_{11} &=   \frac{\mathpzc{g}_n}{2 \mathpzc{g}_i \mathbb{Z}_e} n_i n_e^2 \bar{g}_{\tilde{T}} e^{-2 \lambda} e^{x^*} \int_{x^*}^\infty dx_0 \, e^{-x_0} \cdot x_0 \, \int_{x^*}^{x_0}  x_1 \zeta^{(0)} \left( \sqrt{\lambda x_1} \right) \zeta^{(0)} \left( \sqrt{\lambda x_2} \right) \, \frac{d\sigma^{ion}}{d\upsilon} d \upsilon \\
J^{rec}_{22} &=   \frac{\mathpzc{g}_n}{2 \mathpzc{g}_i \mathbb{Z}_e} n_i n_e^2 \bar{g}_{\tilde{T}} e^{-2 \lambda} e^{x^*} \int_{x^*}^\infty dx_0 \, e^{-x_0} \cdot x_0 \, \int_{x^*}^{x_0}  x_2 \zeta^{(0)} \left( \sqrt{\lambda x_1} \right) \zeta^{(0)} \left( \sqrt{\lambda x_2} \right) \, \frac{d\sigma^{ion}}{d\upsilon} d \upsilon \\
J^{rec}_{01} &=  0\\
J^{rec}_{02} &=  0\\
J^{rec}_{12} &=  \frac{4}{9} \frac{\mathpzc{g}_n}{2 \mathpzc{g}_i \mathbb{Z}_e} n_i n_e^2 \bar{g}_{\tilde{T}} e^{-2 \lambda} e^{x^*} \int_{x^*}^\infty dx_0 \, e^{-x_0} \cdot x_0 \, \int_{x^*}^{x_0} \lambda x_1 x_2 \zeta^{(1)} \left( \sqrt{\lambda x_1} \right) \zeta^{(1)} \left( \sqrt{\lambda x_2} \right) \, \frac{d\sigma^{ion}}{d\upsilon} d \upsilon
\end{align}
\end{subequations}

\end{appendices}

\bibliographystyle{phjcp}

\begin{thebibliography}{10}

\bibitem{oxenius_kinetic_1986}
{J.~Oxenius},
\newblock {\em Kinetic theory of particles and photons},
\newblock Springer, 1986.

\bibitem{bar-shalom_hullac_2001}
{A.~Bar-Shalom}, {M.~Klapisch}, and {J.~Oreg},
\newblock {\em Journal of Quantitative Spectroscopy and Radiative Transfer}
  {\bf 71}, 169 (2001).

\bibitem{gu_flexible_2008}
{M.~Gu},
\newblock {\em Canadian Journal of Physics} {\bf 86}, 675 (2008).

\bibitem{zatsarinny_bsr:_2006}
{O.~Zatsarinny},
\newblock {\em Computer Physics Communications} {\bf 174}, 273 (2006).

\bibitem{jonsson_new_2013}
{P.~Jonsson}, {G.~Gaigalas}, {J.~Bieroń}, {C.~F. Fischer}, and
  {I.~Grant},
\newblock {\em Computer Physics Communications} {\bf 184}, 2197 (2013).

\bibitem{kallman_atomic_2007}
{T.~R. Kallman} and {P.~Palmeri},
\newblock {\em Reviews of Modern Physics} {\bf 79}, 79 (2007).

\bibitem{annaloro_vibrational_2014}
{J.~Annaloro} and {A.~Bultel},
\newblock {\em Physics of Plasmas} {\bf 21}, 123512 (2014).

\bibitem{scott_cretinradiative_2001}
{H.~A. Scott},
\newblock {\em Journal of Quantitative Spectroscopy and Radiative Transfer}
  {\bf 71}, 689 (2001).

\bibitem{chung_flychk:_2005}
{H.-K. Chung}, {M.~Chen}, {W.~Morgan}, {Y.~Ralchenko}, and {\sc
  R.~Lee},
\newblock {\em High Energy Density Physics} {\bf 1}, 3 (2005).

\bibitem{hansen_hybrid_2007}
{S.~Hansen}, {J.~Bauche}, {C.~Bauche-Arnoult}, and {M.~Gu},
\newblock {\em High Energy Density Physics} {\bf 3}, 109 (2007).

\bibitem{kapper_ionizing_2011}
{M.~G. Kapper} and {J.-L. Cambier},
\newblock {\em Journal of Applied Physics} {\bf 109}, 113309 (2011).

\bibitem{panesi_collisional_2013}
{M.~Panesi} and {A.~Lani},
\newblock {\em Physics of Fluids} {\bf 25}, 057101 (2013).

\bibitem{le_complexity_2013}
{H.~P. Le}, {A.~R. Karagozian}, and {J.-L. Cambier},
\newblock {\em Physics of Plasmas} {\bf 20}, 123304 (2013).

\bibitem{munafo_boltzmann_2014}
{A.~Munafo}, {M.~Panesi}, and {T.~E. Magin},
\newblock {\em Physical Review E} {\bf 89} (2014).

\bibitem{guy_consistent_2015}
{A.~Guy}, {A.~Bourdon}, and {M.-Y. Perrin},
\newblock {\em Physics of Plasmas} {\bf 22}, 043507 (2015).

\bibitem{hagelaar_solving_2005}
{G.~J.~M. Hagelaar} and {L.~C. Pitchford},
\newblock {\em Plasma Sources Science and Technology} {\bf 14}, 722 (2005).

\bibitem{longo_monte_2000}
{S.~Longo},
\newblock {\em Plasma Sources Science and Technology} {\bf 9}, 468 (2000).

\bibitem{le_modeling_2015}
{H.~P. Le} and {J.-L. Cambier},
\newblock {\em Physics of Plasmas} {\bf 22}, 093512 (2015).

\bibitem{braginskii_transport_1965}
{S.~I. Braginskii},
\newblock Transport processes in a plasma,
\newblock in {\em Review of plasma physics}, volume~1, pp. 205--311,
  Consultants Bureau, New York, 1965.

\bibitem{burgers_flow_1969}
{J.~M. Burgers},
\newblock {\em Flow equations for composite gases},
\newblock Academic Press, 1969.

\bibitem{horwitz_ion_1973}
{J.~Horwitz} and {P.~Banks},
\newblock {\em Planetary and Space Science} {\bf 21}, 1975 (1973).

\bibitem{barakat_momentum_1981}
{A.~R. Barakat} and {R.~W. Schunk},
\newblock {\em Journal of Physics D: Applied Physics} {\bf 14}, 421 (1981).

\bibitem{benilov_momentum_1996}
{M.~S. Benilov},
\newblock {\em Physics of Plasmas} {\bf 3}, 2805 (1996).

\bibitem{benilov_kinetic_1997}
{M.~S. Benilov},
\newblock {\em Physics of Plasmas} {\bf 4}, 521 (1997).

\bibitem{conde_friction-force_2008}
{L.~Conde}, {L.~F. Ibáñez}, and {J.~Lambás},
\newblock {\em Physical Review E} {\bf 78} (2008).

\bibitem{zeldovich_physics_2002}
{Y.~B. Zeldovich} and {Y.~B. Raizer},
\newblock {\em Physics of shock waves and high-temperature hydrodynamic
  phenomena},
\newblock Dover Publications, 2002.

\bibitem{johnson_cubature_2014}
{S.~G. Johnson},
\newblock Cubature {Package}, 2014.

\bibitem{rosa_magnetohydrodynamic_1968}
{R.~J. Rosa},
\newblock {\em Magnetohydrodynamic {Energy} {Conversion}},
\newblock McGraw-Hill, 1968.

\bibitem{van_der_mullen_excitation_1990}
{J.~van~der Mullen},
\newblock {\em Physics Reports} {\bf 191}, 109 (1990).

\end{thebibliography}

\end{document}